\documentclass[%
reprint,
 amsmath,amssymb,
 aps,
]{revtex4-1}

\usepackage{amsmath} 
\usepackage{amssymb}
\usepackage{eclbkbox}
\usepackage{slashbox}
\usepackage{dcolumn}% Align table columns on decimal point
\usepackage{bm}% bold math
\usepackage{daytime} 
\usepackage{dsfont}
\newcommand{\rset}{\stackrel{\ {\sf R}}{\leftarrow}}
\newcommand{\uset}{\stackrel{\ {\sf U}}{\leftarrow}}
\newcommand{\set}{:=}
\newcommand{\ra}{\rightarrow}

\newcommand{\N}{\mathbb{N}}
\newcommand{\Q}{\mathbb{Q}}
\newcommand{\R}{\mathbb{R}}

\newcommand{\CCC}{{\cal C}}
\newcommand{\DDD}{{\cal D}}

\newcommand{\ZZZ}{{\cal Z}}
\newcommand{\SSS}{{\cal S}}

\newcommand{\MMM}{{\cal M}}

\newcommand{\XXX}{{\cal X}}
\newcommand{\YYY}{{\cal Y}}
\newcommand*{\qed}{\hfill\ensuremath{\square}}
\newcommand{\scl}{{\ \overset{\delta}{\approx}\ }}
\newcommand{\seq}{{\ \overset{\delta(\cdot)}{=}\ }}
\newcommand{\OC}{{\sf OC}}
\newcommand{\SA}{{\sf SA}}

\newcommand{\SSC}{{\sf SSoC}}
\newcommand{\SCC}{{\sf SChC}}
\newcommand{\ECS}{{\sf ECS}}
\newcommand{\SC}{{\sf SCh}}
\newcommand{\ES}{{\sf ES}}

\usepackage{url}

\begin{document} 
%\pagestyle{headings}
%\pagestyle{empty}

% \mainmatter

\title{A Unified Paradigm of Organized Complexity %Measure 
and Semantic Information Theory}

\author{Tatsuaki Okamoto}
\affiliation{NTT \\
3-9-11 Midori-cho, Musashino-shi, Tokyo, 180-8585 Japan\\
%Email: 
{okamoto.tatsuaki@lab.ntt.co.jp}
  %Kyoto University
}

\date{\today}
%August 29, 2015}

\begin{abstract}
One of the most fundamental problems in science is to define {\it
 quantitatively} the complexity of organized matters, i.e., {\it
 organized complexity}.
Although many measures have been proposed toward this aim in previous decades, 
there is no agreed upon definition.
This paper presents a new quantitative definition of organized complexity.
In contrast to existing measures such as the Kolmogorov complexity, logical depth, effective complexity, and statistical complexity, this new definition {\it simultaneously}
captures the three major features of complexity:
computational (similar to logical depth), 
descriptional (similar to the Kolmogorov complexity and effective complexity) 
and distributional (similar to statistical complexity).
In addition, the proposed definition is computable and 
can measure  
both probabilistic and deterministic forms of objects 
in a unified manner.
The proposed definition is based on %(probabilistic) 
circuits rather than Turing machines and $\epsilon$-machines. 
We give several criteria required for organized complexity measures and 
show that the proposed definition 
satisfies all of them for the first time.

We then apply this quantitative definition to formulate a {\it semantic information theory}. We present the first formal definition of a {\it semantic information amount}, which is the core concept
of the semantic information theory, 
that is based only on concretely defined notions.
Previous semantic information theories defined this amount under 
some a priori information
which is not concretely specified. 
We then unveil several fundamental properties in the semantic information theory, e.g., 
a semantic source coding theorem, semantic channel coding theorem, and effectiveness coding theorem. 
Although the semantic information theory has a long history of research going back more than six decades, there has been no study on its relation to organized complexity. This paper offers the first unified paradigm of organized complexity and semantic information theory.

\end{abstract}

% \today,
% \now
%\daytime
%\doublespacing

%\renewcommand{\baselinestretch}{0.95} 
\newtheorem{theorem}{Theorem} 
\newtheorem{definition}{Definition}
\newtheorem{remark}{Remark}

\maketitle

\section{Introduction}

\subsection{Background}
\label{sec:bacgroud}

Around seven decades ago, 
an American scientist, Warren Weaver, classified scientific problems into three classes: 
problems of {\it simplicity}, problems of {\it disorganized} complexity, and
problems of {\it organized} complexity \cite{Weaver48}. 
For example, the classical dynamics can be used to analyze and predict the motion of 
a few ivory balls as they move about on a billiard table. 
This is a typical problem of simplicity. 
Imagine then, a large billiard table with millions of balls rolling over
its surface, colliding with one another and with the side rails. 
Although to be sure the detailed history of one specific ball cannot be
traced, statistical mechanics can analyze and predict the average
motions. This is a typical problem of disorganized complexity. 
Problems of organized complexity, however, deal with features of
organization such as living things, ecosystems, and artificial things. 
Here, cells in a living thing are interrelated into an organic whole in their positions and motions, 
whereas the balls in the above illustration of disorganized complexity
are distributed in a
helter-skelter manner.

In the tradition of Lord Kelvin, 
the quantitative definition of complexity is the most fundamental and
important notion in problems of
complexity.

\vspace{5pt}

{\it ``I often say that when you can measure what you
are speaking about, and express it in numbers, you
know something about it; but when you cannot
measure it, when you cannot express it in numbers,
your knowledge is of a meagre and unsatisfactory
kind; it may be the beginning of knowledge, but you
have scarcely in your thoughts advanced to the
state of science, whatever the matter may be.''}

\begin{flushright}
{\it Lord Kelvin, 1883}
\end{flushright}

\vspace{5pt}

The quantitative definition of 
{\it disorganized} complexity 
of physical systems 
has been established to be {\it entropy}, which is defined in 
thermodynamics and statistical mechanics. 
In a similar manner, disorganized complexity 
of information sources (distributions) 
can be quantitatively defined 
by Shannon entropy \cite{Shannon48}. 

In contrast, there is no agreed upon quantitative definition of 
{\it organized} complexity.
The difficulty comes from the notion that organized complexity
could be greatly dependent on our senses or that the objects 
of organized complexity like living things, ecosystems, and artificial things may be 
recognized only by intelligent organisms like human beings,
that is to say, it is vastly different from
the measures of disorganized complexity 
such as entropy and Shannon entropy 
which simply quantify the randomness of the objects.

We may therefore wonder whether such
sensory and vague things can be rigorously defined 
in a unified manner covering 
various living things to artificial things.
Many investigations nonetheless have been pursued toward this aim in the last
decades, e.g., {\it logical depth} by Bennett \cite{Bennett88}, 
{\it effective complexity} by Gel-Man \cite{Gell-Mann95,GelLlo96,GelLlo04},
{\it thermodynamic depth} by Lloyd and Pagels \cite{LloPag88},
{\it effective measure complexity} by Grassberger \cite{Grassberger89},
and {\it statistical complexity} by Crutchfield et al. \cite{CruYou89,Cruchfield94,CruSha99,ShaCru01},
although no existing measure has been agreed on 
in the field.
\cite{LinDavRus13,LayLamWie13}.
In Section \ref{sec:survey}, we explain our understanding on why no existing
measure 
%can be agreed upon satisfactorily.
is satisfactory to be agreed upon.

The quantitative definition of complexity of an object is 
essentially related to the amount of information that the object possesses.
For example, Shannon entropy, which is the quantitative definition of
disorganized complexity of an information source,
was introduced to define the amount of information of a source  
in the sense of Shannon's information theory
\cite{Shannon48}.

One year
after Shannon introduced his information theory (and Weaver published the aforementioned article \cite{Weaver48}), Weaver proposed that
there are three levels of information and communication problems \cite{Weaver49}:

\vspace{5pt}

\noindent
- {Level A:} \ How accurately can the symbols of
communication be transmitted? (The technical problem.)

\vspace{3pt}
\noindent
- {Level B:} \ How precisely do the transmitted symbols
convey the desired meaning? (The semantic
problem.)

\vspace{3pt}
\noindent
- {Level C:} \ How effectively does the received meaning
affect conduct in the desired way? (The effectiveness
problem.)

\vspace{5pt}

Interestingly, 
the classes of scientific problems classified by Weaver \cite{Weaver48}
are closely related to the above-mentioned three levels of information problems.
The {\it disorganized} complexity measures, e.g., Shannon entropy,
are related to the Level A problem, {\it technical or syntactic} problem, e.g., Shannon information theory,
and the {\it organized} complexity measures should be related to the Level B and C
problems, i.e., {\it semantic and effectiveness} problems.

There have been many studies on the Level B 
(semantic) problem spanning more than six decades 
in terms of the semantic information theory 
\cite{BaoBasDeaParSwaLel11,Bacchus88,CarBar52,D'alfonso11,Flordi04,Flordi09,
JubSud08i,JubSud08ii,KohSch09,Langel09,NafAle09,Nillsson86,Sommagura09,
WilKal05}, 
but no existing work has been 
recognized as a standard theory. 
In Section \ref{sec:existing-semantic-theories},
we show a fundamental problem common among the existing works for Level B
that forms the basis of our understanding of why no existing work 
could be a standard theory.
Roughly, 
all existing work assumes some a priori information 
which is not concretely specified, 
where informal observation might be possible but 
no formal result could be achieved rigorously without 
concrete specification of such a priori information.
In addition, no observation has been presented in literature on the relation between the semantic (Level B) problem and the organized complexity. 
To the best of our knowledge,
no study has been conducted seriously on the Level C (effectiveness) problem.

\subsection{Contribution}

This paper presents a new quantitative definition of organized complexity.
In contrast to existing measures such as Kolmogorov complexity, logical depth, effective complexity and statistical complexity, this new definition simultaneously captures the three major features of complexity:
computational feature (similar to logical depth), 
descriptional feature (similar to Kolmogorov complexity and effective complexity) 
and distributional feature (similar to statistical complexity).
In addition, the proposed definition is computable and 
can measure both probabilistic and deterministic forms of objects 
in a unified manner.
The proposed definition is based on 
%the shortest size of a (stochastic finite-state automaton form of) 
{\it circuits} 
\cite{Sipser96,Vollmer99} %oc-circuit, 
rather than %properties on 
Turing machines \cite{Sipser96,LiVit97,Bennett88,Gell-Mann95,GelLlo96,GelLlo04}
and $\epsilon$-machines \cite{CruYou89,Cruchfield94,CruSha99,ShaCru01}.
Our new measure is given by the shortest size of a stochastic finite-state automaton form of circuit, {\it oc-circuit}, for simulating the object.  
Here note that, given an object, the shortest size of an oc-circuit to simulate the object is computable and that  
the size of an oc-circuit can capture
the computational, descriptional 
and distributional features of complexity of the object.
%that the oc-circuit simulates. 
We give several criteria required for organized complexity measures and 
show that the proposed definition is the first that 
satisfies all of the requirements. 

We then present the first semantic information theory for the Level B (semantic) problem that 
overcomes the fundamental problem common among all previous works. 
That is, the proposed semantic information theory
is constructed 
only on concretely defined notions. 
This theory is based on the proposed organized complexity measure.
We then unveil several fundamental properties in the semantic information theory, e.g., 
a semantic source coding theorem and semantic channel coding theorem.
Moreover, this paper, for the first time, develops a theory for the effectiveness (Level C) problem,
which is also constructed on our organized complexity measure.
In other words, we 
clarify the relationship of
organized complexity with the semantic and effectiveness (Level B 
and C) problems of information and communication.

Thus, this paper presents the first unified paradigm for the organized complexity
and the semantic information theory that covers
the semantic and effectiveness problems.

\subsection{Notations}
\label{sec:notation}

The sets of natural, rational, and real numbers
are denoted by $\N$, $\Q$, and $\R$, 
respectively. 
The set of $n$-bit strings is denoted by $\{0,1\}^n$ ($n \in \N$),
$\{0,1\}^* \set \cup_{n \in \N} \{0,1\}^n$,
and
the null string (0-bit string) is denoted by $\lambda$. 
When $x \in \{0,1\}^*$, $|x|$ denotes the bit length of $x$. 
When $a, b \in \R$, 
$[a, b]$ denotes set $\{x \mid x \in \R, \   
a\leq x \leq b \} \subset \R$. 
When $x\in\R$,   
$\lceil x \rceil$ denotes the smallest integer
greater than or equal to $x$.   

When $x$ is a variable and $y$ is a value or %equation specification, 
$x \set y$ denotes that 
$x$ is substituted or defined by $y$.
A probability distribution over $\{0,1\}^n$ is 
$\{ (a, p_a) \mid a \in \{0,1\}^n, 
p_a \in [0,1], 
%p_a \in \R, 0\leq p_a \leq 1, 
\sum_{a\in \{0,1\}^n}p_a=1 \}$. 
When $A$ is a probability distribution,
or the source (machinery) of the distribution,
$a \rset A$ denotes that 
element $a \in \{0,1\}^n$ is randomly selected from $A$
according to its probability distribution.
When $A$ is a set, $a \uset A$ denotes that 
$a$ is randomly selected from $A$ with a uniform distribution.
   
When $X$ and $Y$ are two distributions,  
the statistical distance of $X$ and $Y$,
${\sf SD}(X,Y)$, is defined by
$\frac{1}{2}\cdot \sum_{\alpha \in \{0,1\}^*} | \Pr[\alpha \rset X]
- \Pr[\alpha \rset Y] |$,
and
$X \overset{\delta}{\approx} Y$ 
denotes that ${\sf SD}(X,Y)$ 
is bounded by $\delta$.
Then we say $X$ and $Y$ are 
statistically $\delta$-close.

When $Y$ is a distribution,
$(Y)_n$ denotes the $n$-bit restriction of $Y$,
i.e., $(Y)_n$ is a distribution over $\{0,1\}^n$ 
and
$\Pr[y \rset (Y)_n] = \sum_{z \in \{0,1\}^*} \Pr[(y,z) \rset Y] $.

When $S$ is a set, ${\#}S$ denotes the number of elements of $S$.

\section{Organized Complexity Measure}

\subsection{Objects}

The existing complexity measures can be categorized in two classes.
One is a class of measures whose objects are {\it deterministic 
strings}, 
and the objects of the other class of measures 
are {\it probability distributions}.
As for the traditional 
complexity measures, 
the Kolmogorov complexity is categorized in the former and (Shannon) entropy is in the latter.     
Among the above-mentioned organized complexity measures, 
logical depth and effective complexity are in the former, and
thermodynamic depth, effective measure complexity, 
and statistical depth are in the latter.

Which is more appropriate as the objects of organized complexity?

The objects of complexity are everything around us, 
stars and galaxies in space, living things, ecosystems, 
artificial things, and human societies.
The existence of everything can be recognized by us only through observations. 
For example, the existence of many things are observed through devises such as the telescope, microscope,  
various observation apparatus and electronic devices. 
We can take things around us directly into our hands and sense them, but they are also recognized by our brains as electronic nerve signals 
transmitted from the sensors of our five senses through the nervous system.
That is, all objects of complexity are recognized as the result of observations by  
various apparatus and devices including the human sensors of our five senses.
  
Since the micro world is governed by quantum mechanics, 
observed values are determined in a probabilistic
manner. This is because observed values (data) obtained when observing 
micro phenomena (quantum states) in quantum mechanics are 
randomly selected according to a certain probability distribution
corresponding to a quantum state (e.g., entangled superposition). 

How then, 
are observed values in macro phenomena? 
For example, 
if some sort of radio signals are received and measured, they would
almost certainly be accompanied by noise. There are various reasons 
why noise becomes mixed in with signals, and 
one of them  
is thought to be the probabilistic phenomena of electrons,
thermal noise. 
Similarly, various types of noise will be present 
in the data obtained when observing distant astronomical bodies. 
This can be caused by the path taken by the light (such as through the atmosphere) and by factors associated with the observation equipment.

Even in the case of deterministic physical phenomena, chaos theory 
states that fluctuations in initial conditions can lead to diverse 
types of phenomena that behave similar to those of random systems. In short, 
even a deterministic system can appear to be a quasi-probabilistic
system. But even a system of this type can become a true (non-quasi) 
probabilistic system if initial conditions fluctuate 
due to some noise, e.g., thermal noise.
There are also many cases in which a quasi-probabilistic system 
associated with chaos cannot be distinguished from a true probabilistic 
system depending on the precision of the observation equipment. 
Here, even quasi-probabilistic systems may be treated as 
true probabilistic systems.

Thus, when attempting to give a quantitative definition of the complexity of observed 
data, the source of those observed data would be a probability 
distribution and the observed data themselves would be values randomly 
selected according to that distribution.
If we now consider the complexity of a phenomenon observed 
using certain observation equipment, the object of this complexity 
should not be the observed data selected by chance from the source 
but rather the probability distribution itself corresponding to 
the source of the observed data.

It is known that some parts of genome patterns
appear randomly distributed over a collection of many samples (over generations).
Here, %under ergodicity, 
we can suppose a source (probability distribution)
of genome patterns, from which each genome pattern is randomly selected.
Also in this case,
the object of complexity 
should not be each individual genome pattern
but rather the source (probability distribution) of the genome patterns.

Therefore, hereafter in this paper,
we consider that an object of complexity is a {\it probability distribution}.
Here note that a deterministic string can be considered to be a very
special case of a probability distribution
(where only a value occurs with probability 1 and the others with 0).

How can we determine a source or probability distribution from observed data?
It has been studied as the {\it model selection theory} in statistics and information
theory, e.g., AIC (Akaike's information criterion) by Akaike \cite{Akaike73}
and MDL (minimum description length) by Rissanen \cite{Rissanen84,Rissanen89},
given a collection of data,
to find the most likely and succinct model (source, i.e., probability distribution) of the data. 
In this paper, however, it is outside the scope, i.e., 
we do not consider how to find such a source from a collection of observed data (through the model selection theory). 
We here suppose that 
a source (probability distribution) 
is given as an object of organized complexity, 
and focus on how to define quantitatively the complexity of such a given source.

There are roughly two types of observed data,
one type is data observed at a point in time and the other is time series data. 
Genome pattern data are an example of the former, and
data obtained from an observation apparatus for a certain time period
are an example of the latter.
In any case, 
without loss of generality,
we here assume that observed data $x$ are bounded and 
expressed in binary form, i.e.,
$x \in \{0,1\}^n$ for some $n \in \mathbb{N}$,
since any physically observed data have only finite precision (no infinite precision).
Then the {\it source} of the observed data, $X$,
which is an object of organized complexity in this paper,
is a probability distribution over $\{0,1\}^n$ 
for some $n \in \mathbb{N}$ such that
$X := \{ (x, p_x) \mid x \in \{0,1\}^n, \ 0\leq p_x \leq 1, \ 
\sum_{x \in \{0,1\}^n} p_x = 1 \}$.

\subsection{Criteria}
\label{sec:criteria}

We describe our attempt to define quantitatively the organized complexity.
To begin with, let us consider the following example. We give a
chimpanzee a computer keyboard and prompt the chimpanzee to hit the keys 
freely resulting in the output of a string of characters. Let us assume 
an output of 1000 alphabetical characters. At the same time, we select 
a string of 1000 characters from one of Shakespeare's plays. Naturally, 
the character string input by the chimpanzee is gibberish possessing no 
meaning, which undoubtedly makes it easy for us to distinguish that
string from a portion of a Shakespearean play. 

Is there a way, however, to construct a mathematical formulation of 
the difference between these two strings that we ourselves can easily 
tell apart? Why is it so easy for us to make a distinction between these 
two strings? The answer is likely that the chimpanzee's string is simply random (or disorganized) and meaningless to us
while Shakespeare's string is highly organized and meaningful. In short, if we can mathematically define the amount of organized complexity
(or meaningful information), 
we should be able to make 
a distinction between these two strings.

What then are the sources of the observed data,
the chimpanzee's string and Shakespeare's string.
Let us return to the source of the chimpanzee's string creation 
without thinking of it as simply a deterministic string. 
Here, for the sake of simplicity, 
we suppose that the scattered hitting of keys by the chimpanzee is the 
same as a random selection of hit keys. At this time, the 
source of the chimpanzee's string is the probability distribution in 
which any particular 1000-character string can be randomly selected 
from all possible 1000-character strings with equal probability. 

What, then, would be the source of Shakespeare's string? We can surmise
that, when Shakespeare wrote down this particular 1000-character string, 
a variety of expressions within his head would have been candidates 
for use, and that the 1000-character string used in the play would have 
finally been selected from those candidates with a certain probability. 
The candidates selected must certainly be connected by complex semantic 
relationships possessed by English words. Accordingly, the source of 
Shakespeare's string must be the complex probability distribution of 
candidate expressions connected by complex semantic relationships.
For example (Case 1), candidate expression 1 has the probability of 0.017, 
candidate expression 2 has the probability of 
0.105, ..., candidate expression 327 has the probability of 0.053 and the other expressions have the probability of 0,
i.e., hundreds of candidate expressions occurred in 
his head consciously or unconsciously and finally one of them was randomly chosen according to the distribution.
As more simplified cases,
Case 2 is where candidate expression 1 has the probability of 2/7, candidate expression 2 has the probability of 5/7 
and the others have the probability of 0, i.e., only two candidate expressions occurred in 
his head and finally one of them was randomly chosen. 
Case 3 is where only a single expression has the probability of 1 and the others have the probability of  0, i.e., a deterministic string case; he selected the expression without hesitation.  
These expressions as well as the distributions should be highly 
organized and structured with complex semantic relationships.

Considering the above-mentioned observation,
we give the following criteria for formulating the organized
complexity measures.

\begin{enumerate}
\item
The objects should be probability distributions. 
In addition, deterministic strings (as a special case of distributions) 
and more general distributions should be treated in a unified manner, 
e.g., the complexity of Cases 1, 2 and 3 for the source of Shakespeare's string
should be measured in a unified manner. 

\item
Simple (or very regular) objects, which are treated as ``problems of
simplicity'' based on Weaver's classification, should have low organized
complexity.

\item
Simply random objects, which are treated as ``problems of 
disorganized complexity'' by Weaver, e.g., the source of the chimpanzee's 
string, should have low organized complexity.  

\item 
Highly organized objects, which are treated as ``problems of 
organized complexity'' by Weaver, 
e.g., Cases 1, 2 and 3 for the source of Shakespeare's string,
should have high organized complexity.  

\item
The organized complexity of an object should be computable (or recursive in
computation theory).

\end{enumerate}

\subsection{Existing Complexity Measures}
\label{sec:survey}

Using these criteria,
we now survey the typical quantitative definitions of organized
complexity in literature.

\vspace{7pt}
\noindent
{\bf Objects are ``deterministic strings''}

\begin{itemize}

\item{\bf Kolmogorov complexity}

The notion of the Kolmogorov complexity was independently proposed
by Solomonoff, 
Kolmogorov, and Chaitin 
\cite{Solomonoff64,Kolmogorov65,Chaitin69,LiVit97}.

Roughly,
the Kolmogorov complexity of string $x$
is the size of the shortest program (on a computer) to produce string $x$.

More precisely, 
let $U$ be a reference universal prefix (Turing) machine
(see \cite{LiVit97} for the reference universal prefix machine).
Then, 
the Kolmogorov complexity, $K(x)$, of string $x \in \{0,1\}^*$ is
defined by 
$$
K(x) = \min\{ |z| \mid  U(z) = x, z \in \{0,1\}^* \}.
$$

In light of the above-mentioned criteria,
the Kolmogorov complexity has the following properties.

\begin{enumerate}
\item
The objects are only deterministic strings (Bad).

\item
Simple (or very regular) objects have low Kolmogorov complexity (Good).

\item
Simply random objects, deterministic $n$-bit strings, that are 
uniformly and randomly chosen from $\{0,1\}^n$   
have high Kolmogorov complexity (Bad).

\item 
Highly organized objects may have between high and low logical depth (Bad),
since some highly organized complex objects that
are generated from small strings 
through very long running-time and complex computations 
may have low Kolmogorov complexity. 
In other words, 
some organized complexities may be characterized 
in a dynamic manner as logical depth rather than 
in a static manner as Kolmogorov complexity. 

\item
The Kolmogorov complexity of an object (string) is not computable (Bad).

\end{enumerate}

\item {\bf Logical depth} 

Bennett \cite{Bennett88} introduced {\it logical depth} with the intuition that
complex objects are those whose most plausible explanations describe
long causal processes.
To formalize the intuition, Bennett employs the methodology of
algorithmic information theory, the Kolmogorov complexity.

The logical depth of a deterministic string, 
Bennett's definition for measuring organized complexity,
is dependent on the running time
of the programs that produce the string and whose length is relatively close to
the minimum in a sense. 

More precisely,
the logical depth of string $x$ at significance level 
$\epsilon \set 2^{-b}$ \cite{LiVit97} is
$$\min \{ t \mid m_t(x)/m(x)) \ge \epsilon\},$$
where we define $m_t(x)$ and $m(x)$ by
$$ m_t(x) := \sum_{U^t(p)=x} 2^{-l(p)},$$ \ \ \ 
$$m(x) := \sum_{U(p)=x} 2^{-l(p)}.$$
Here, $U$ is the reference universal prefix (Turing) machine 
(for the Kolmogorov complexity) and 
$U^t$ is a specific class of $U$ whose running time is
bounded by $t$ steps.
$l(p)$ is the length of program $p$.

In light of the above-mentioned criteria,
the logical depth has the following properties.

\begin{enumerate}
\item
The objects are only deterministic strings (Bad).

\item
Simple (or very regular) objects have low logical depth (Good).

\item
Simply random objects have low logical depth (Good).

\item 
Highly organized objects may have between high and low logical depth (Bad),
since some highly organized complex objects may have low logical depth 
with relatively high Kolmogorov complexity, where the core part of the organized complexity is 
due to the Kolmogorov complexity. 
In other words, 
some organized complexities may be characterized 
in a static manner as Kolmogorov complexity rather than
in a dynamic manner as logical depth,
(where $m_t(x)/m(x) \ge \epsilon$ is required but
the logical depth is not dependent on the value of $m(x)$ itself
(roughly, $-\log m(x)$ is close to the Kolmogorov complexity of $x$)).

\item
The logical depth of an object (string) is not computable, 
since it is based on the Kolmogorov complexity or universal Turing
machines (Bad).

\end{enumerate}

\item {\bf Effective complexity}
 
Effective complexity \cite{Gell-Mann95,GelLlo96} was introduced by Gell-Mann, and is 
based on the Kolmogorov complexity. To define the complexity of an object,
Gell-Mann considers the shortest description of the distribution in which
the object is embedded as a typical member. 
Here, `typical' means that the negative logarithm
of its probability is approximately equal to the entropy of the distribution. 

That is, the effective complexity of string $x$ is
$$\min\{K(E) \mid - \log {\rm Pr}_E(x) \approx H(E) \}, $$
where $K(E)$ is the Kolmogorov complexity of distribution $E$,
i.e.,  the length of the shortest program to list all members, $r$, of $E$
together with their probabilities, $\Pr_E(r)$, and $H(E)$ is the (Shannon) entropy of $E$.

In light of the above-mentioned criteria,
the effective complexity has the following properties.

\begin{enumerate}
\item
The objects are only deterministic strings (Bad).
Technically, however, we can consider distribution $E$ to be an object of the effective complexity.

\item
Simple (or very regular) objects have low effective complexity (Good).

\item
Simply random objects have low effective complexity (Good).

\item 
Highly organized objects may have between high and low effective complexity  (Bad),
since some highly organized complex objects may have low effective complexity 
with very high computational complexity of the universal machine to generate $E$, 
where the core part of the organized complexity is due to the computational complexity in a dynamic manner
(e.g., high logical depth of $E$).   

\item
The effective complexity of an object (string) is not computable, 
since it is based on the Kolmogorov complexity or universal Turing
machines (Bad).

\end{enumerate}

\end{itemize}

\vspace{7pt}
\noindent
{\bf Objects are ``probability distributions''}

\begin{itemize}

\item {\bf Thermodynamic depth} 

The {\it thermodynamic depth} was introduced by Lloyd and Pagels \cite{LloPag88} and shares
some informal motivation with logical depth,
where complexity is considered a property of the evolution of an object.

We now assume the set of histories or trajectories that result in 
object (distribution) $S_0$. A trajectory is an ordered set of macroscopic states (distributions)
$S_{-L-1}, .., S_{-1}, S_0$. 
The thermodynamic depth of object $S_0$ is 
$$H(S_{-L+1}, .., S_{-1} \mid S_0),$$
where $H(A,..,B \mid C)$ is the conditional entropy of combined distribution $(A,..,B)$
with condition $C$.

One of the major problems with this notion is that it is not defined 
how long the trajectories (what value of $L$) should be.
Moreover, it is impossible to specify formally the trajectories, 
given an object,
since there is no description on how to select macroscopic states
in \cite{LloPag88}.
If there are thousands of possible sets of macroscopic states, we would have thousands of different definitions of the thermodynamic depth.

Another
fundamental problem with this measure is that
in order to measure the complexity of an object $S_0$,
a set of macroscopic states 
$\SSS \set \{S_{-L+1}, .., S_{-1}\}$
whose complexity is comparable to or more than that of $S_0$ should be established beforehand.
Hence, 
if $\SSS$ is fixed, or
the thermodynamic depth of $\SSS$ is concretely defined, it cannot measure the complexity of an object whose complexity is more than that of $\SSS$.
That is, any concrete definition of this notion can measure only a restricted subset of objects, i.e., any concrete and generic definition is impossible in thermodynamic depth.    
It should be a fundamental problem with this concept.

As a result, it is difficult to define rigorously the thermodynamic depth
and to characterize the definition. 

Note that we have the same criticisms for 
the existing semantic information theories that are described in Section 
\ref{sec:existing-semantic-theories}.

%%%%% 20160629

\item {\bf Effective measure complexity}

The {\it effective measure complexity} was introduced by Grassberger 
\cite{Grassberger89} 
and measures the average amount by which the uncertainty of a symbol in a string   
decreases due to the knowledge of previous symbols.

For distribution $X^N$ over $\{0,1\}^N$ $(N \in\N)$, 
$H(X^N)$ is the Shannon entropy of $X^N$.
Let $h_N \set  H(X^{N+1}) - H(X^N)$, and $h \set \lim_{N\rightarrow \infty} h_N$.
The effective measure complexity of $\{X^N \}_{N \in \N}$ is
$$ \sum_{N=0}^{\infty} (h_N - h)$$
This difference quantifies the perceived randomness which, after further observation, is discovered
to be order \cite{LayLamWie13}.

In light of the above-mentioned criteria,
the effective measure complexity has the following properties:

\begin{enumerate}
\item
The objects are only probability distributions, 
and deterministic strings are outside the scope
     of this measure (the complexity is 0
for any deterministic string)
 (Bad).

\item
Simple (or very regular) objects have 
low effective measure complexity (Good).

\item
Simply random objects have low 
effective measure complexity (Good).

\item 
Highly organized objects may have between high and low 
effective measure complexity  (Bad),
since some highly organized complex objects (distributions)
may have low effective measure complexity 
with very high Kolmogorov complexity or computational complexity of the universal machine to
      generate them, 
where the core of the organized complexity is due to
some Kolmogorov complexity or the computational complexity.
In other words, the effective measure complexity cares 
only about distributions but not the
computational features that logical depth and effective complexity
care about.   

\item
The effective measure complexity of an object (distribution) is computable, 
since it is not based on any Turing machine (Good).

\end{enumerate}

\item {\bf Statistical complexity} 

The {\it statistical complexity} was introduced by 
Crutchffeld and Young \cite{CruYou89}.
Here,
to define the complexity, the set of causal states 
$S$ and the probabilistic transitions 
between them are modeled in the so-called $\epsilon$-machine, 
which produces a stationary distribution of causal states, $D_S$.
The mathematical structure of the $\epsilon$-machine is  
a stochastic finite-state automaton or hidden Markov model. 

Let $S_i$ for $i=1,..,k$ be causal states,
$S := \{S_1, ..., S_k\}$ and
$T_{ij}$ be the probability of a transition from state $S_i$ to state $S_j$,
i.e., $T_{ij} := \Pr[S_j \mid S_i]$.
Each transition from $S_i$ to $S_j$ is associated with an output symbol, 
$\sigma_{ij}$, (e.g., $\sigma_{ij} \in \{0,1\}$).  
Then, $\Pr[S_i]$, the probability that $S_i$ occurs in the infinite run of
the $\epsilon$-machine, is given by the eigenvector of matrix $T := (T_{ij})$, since 
$\sum_{i=1}^k \Pr[S_i] \cdot T_{ij} = \Pr[S_j]$, i.e., 
$(\Pr[S_1], ..., \Pr[S_k]) \cdot T = (\Pr[S_1], ..., \Pr[S_k])$.
Hence, the machine produces a stationary distribution of states, $D_S$.   
The output of the $\epsilon$-machine is the infinite sequence of 
$\sigma_{ij}$ induced by the infinite sequence of the transition of states.  
That is, $\epsilon$-machine outputs a distribution, $\Sigma_S$, over $\{0,1\}^\infty$, 
induced by $D_S$.  

The statistical complexity of object $X$ (distribution), 
denoted $C_1$, is the minimum value of the Shannon entropy of $D_S$, $H_1(D_S)$, 
when $\Sigma_S=X$:
$$ C_1 \set \min\{ H_1(D_S) \mid \Sigma_S = X \}. $$

A more generalized notion, $C_\alpha$ ($0\leq \alpha \leq \infty$), is defined by the Reny entropy of $D_S$ in place of the Shannon entropy, i.e., 
$$ C_\alpha \set \min\{ H_\alpha(D_S) \mid \Sigma_S = X \}, $$ 
where
$C_1$ is the case where $\alpha\set 1$ as $H_1$ is the Shannon entropy, and $C_0 \set  
\min\{ \log{{\#}S} \mid \Sigma_S = X \}$ ($\alpha\set0$) 
(${\#}S$ is the number of elements of set $S$) 
(For $\alpha\set\infty$, $H_\infty$ is the mini-entropy).

In light of the above-mentioned criteria,
the statistical complexity has the following properties:

\begin{enumerate}
\item
Statistical complexity $C_1$ can measure only  
probability distributions as objects, since $C_1 =0$
for any deterministic string.
Complexity $C_0$ cannot measure the deterministic strings well either,
since the organized complexity may be around $|S|$ for 
high Kolmogorov complexity or high logical depth
deterministic strings but $C_0 = \log{|S|}$. 
Moreover, $C_0$ cannot capture the distribution 
of the $\epsilon$-machine,
since it only depends on the number of vertexes of causal states.
That is, 
none of $C_\alpha$ with a value of $\alpha$ 
($0\leq \alpha \leq \infty$) can measure
probability distributions and deterministic strings 
in a unified manner (Bad).

\item
Simple (or very regular) objects have low statistical complexity (Good).

\item
Simply random objects have low statistical complexity (Good).

\item 
As for the standard definition of statistical complexity,
i.e., $\alpha\set1$ or the Shannon entropy,
highly organized objects may have between high and low statistical complexity (Bad),
since 
(1) some highly organized complex objects (almost deterministic strings)
may have low statistical complexity, almost zero, 
where the core of the organized complexity is due to
the complexity of the almost deterministic data part,
and
(2) 
some highly organized complex objects (distributions)
have relatively low statistical complexity 
with highly complex output mapping $\{\sigma_{ij}\}$ 
of $\epsilon$-machines,
where the core of the organized complexity is due to
the complexity of $\{\sigma_{ij}\}$ of $\epsilon$-machines
(statistical complexity $C_\alpha$ depends on only $D_S$
but is independent of the complexity of output mapping $\{\sigma_{ij}\}$).
   
See Remark \ref{rmk:statistical complexity} 
for more precise observation.

\item
The statistical complexity of an object (distribution/strings) is computable, 
since it is not based on any Turing machines (Good).

\end{enumerate}
 
\end{itemize}

In summary,
the existing measures have the following drawbacks.
\begin{itemize}
\item
Every existing complexity measure focuses on 
a single feature of complexity, for example,
logical depth focuses on the computational complexity feature, 
the Kolmogorov complexity and effective complexity focus on 
the descriptional complexity feature,
and statistical complexity focuses on the
distributional complexity feature,
but no existing measure captures all of them simultaneously.

\item
Every existing complexity measure can treat either 
a probabilistic or deterministic form of an object,
but no measure can cover both in a unified manner.
 
\item
Some of the measures, the Kolmogorov complexity, 
effective complexity and logical depth, that are based on Turing machines are 
not computable. 

\end{itemize}

\subsection{Proposed Organized Complexity Measure}
\label{sec:proposed}

We now propose a new quantitative definition of organized complexity.
Roughly speaking,
the proposed quantitative definition 
is given by the shortest description size of a (stochastic finite-state automaton form of) {\it circuit} \cite{Sipser96,Vollmer99} that simulates 
an object (probability distribution).

In the existing complexity measures surveyed in Section \ref{sec:survey},
some computing machineries are employed. 
In the logical depth and effective complexity, 
universal Turing machines are employed, 
which cause the uncomputability of their measures.
In the effective measure complexity, no computational machinery is used;
hence, it cannot capture the computational and descriptional features of organized complexities, which logical depth and effective complexity capture, respectively.
The statistical complexity employs
$\epsilon$-machines, whose mathematical model is a stochastic finite-state automaton or hidden Markov model; hence it captures the distributional features of organized complexities 
but not the computational, descriptional, %features 
%the complexities of the output mapping of the $\epsilon$-machine 
and deterministic-object features. %cannot be captured by this complexity measure,
See Remark \ref{rmk:statistical complexity} for more details.        

In the place of universal Turing machines and 
$\epsilon$-machines,
%(stochastic) finite-state automata,
we employ another class of machinery, 
a stochastic finite-state automaton form of 
{\it circuit}, {\it oc-circuit}. Our new measure is given by
the {\it shortest} description size of an oc-circuit for simulating the object.
That is, Occam's razor plays a key role in our definition. 
The advantage of using circuits is that
it can capture the computational 
and distributional features of complexity 
as the size of a circuit
as well as the descriptional 
features of complexity as the input size of a circuit.
Moreover, the shortest description size of an oc-circuit for simulating an object 
is computable (Theorem \ref{thm:computable}), in contrast to that in which the shortest program size on a Turing
machine is uncomputable \cite{LiVit97}. 
Our approach is more general than the approach by $\epsilon$-machines 
in the statistical complexity, since
our oc-circuit model can simulate any $\epsilon$-machine as a special case (Theorem \ref{thm:sim-epsilon}).

The major difference between circuits and Turing machines is that a single (universal) Turing machine can compute any size of input, while a single circuit can compute a fixed size of input.
In spite of the difference, 
any bounded time computation of a Turing machine can be computed by a bounded size of a circuit \cite{Sipser96,Vollmer99}. 
Hence, the proposed complexity measure based on circuits captures general computational features in complexity.       
In addition, the finiteness of 
each circuit yields the computability of our measure, in contrast to the uncomputability of Turing machine based measures such as
logical depth and effective complexity
as well as the Kolmogorov complexity.

We now define a new measure of organized complexity.
First, we define our computation model, {\it oc-circuit}.
 
\begin{definition}(OC-Circuit)
\label{def:oc-circuit}
 \ \ 
Let circuit $C$ with $N$ input bits and $L$ output bits be a directed acyclic graph in which 
every vertex is either an input gate of in-degree 0 labeled by one of 
the $N$ input bits, or one of the basis of gates $B$ := \{AND, OR, NOT\}.
Among them, $L$ gates are designated as the output gates.
That is, circuit $C$ actualizes a Boolean function: 
$\{0,1\}^N \rightarrow \{0,1\}^L$.

Let $s_{i} \in \{0,1\}^{N_s}$ be a state at step $i$ ($i \in \mathbb{N}$), 
$u \in \{0,1\}^{N_u}$ be an a priori input (universe),  
$m_i \in \{0,1\}^{N_m}$ be an input at step $i$,
$r_i \uset \{0,1\}^{N_r}$ be random bits 
at step $i$,
and $N := N_u + N_s + N_m + N_r$. Then,
$$
(s_{i+1}, y_{i}) \longleftarrow 
\fbox{$C(u, \ \cdot \ )$}
\longleftarrow (s_{i}, m_i, r_i), \ \ i=1,2,..., K, 
$$
i.e., \  
$ (s_{i+1}, y_{i}) := C(u, s_{i}, m_i, r_i),$ 
where 
$y_{i} \in \{0,1\}^{L_y}$, $N_m \leq L_y$ is the output of $C$ at step $i$,
and $L := N_s + L_y $.
Let $V$ be the number of vertexes of $C$.

Let $\tilde{C} \set
((w_{ij})_{i=1,..,V; j=1,..,V},
(\ell_1, .., \ell_V), (o_1, .., o_L) )$
be a canonical description of $C$,
where
$(w_{ij})_{i=1,..,V; j=1,..,V}$
is the adjacent matrix of directed graph $C$, 
i.e., $w_{ij} := 1$ iff there is an edge from
vertex $i$ to vertex $j$, and $w_{ij} := 0$ otherwise, 
$\ell_i$ ($i=1,..,V$) is the label of the $i$-th vertex,  
$\ell_i \in \{1,..,N, \mbox{AND, OR, NOT} \}$, 
i.e., each vertex $i$ is labeled by $\ell_i$,
and
$o_i \in \{1,..,V\}$ is the vertex designated to the $i$'s output,
i.e., $(o_1, .., o_L)$ is the sequence of output gates.
Hereafter, we abuse the notation of $C$ to denote $\tilde{C}$, the canonical description of $C$.

Let ${\cal C} \set ( \overline{C}, 
u, n, \vec{m})$ be an ``oc-circuit'',
and $Y$ be the output of ${\cal C}$, 
where 
$\overline{C} \set (C, N_u, N_s, N_m, N_r, L_y, s_1),$
$K \set \lceil n/L_y \rceil$,
$\vec{m} \set (m_1,$ $..,$ $m_K)$,
$Y 
\set (y_1,...,y_K)_n$
(see Section \ref{sec:notation} for the notation of $(...)_n$).

The output, $Y$, of $\CCC$ 
can be expressed by 
$Y \rset \CCC$,
i.e.,
$Y \rset 
(\overline{C}, u, n,\vec{m}_n)$,
where the probability of distribution $Y$ is taken over the randomness of $r_i \uset \{0,1\}^{N_r}$ ($i=1,..,K$).

Then, $\overline{C}$,  
$u$, and $\vec{m}$ 
are called the ``logic,'' ``universe,'' and
``semantics'' of oc-circuit $\CCC$,  
respectively.

\end{definition}

\begin{remark}
\label{rmk:random}
Circuit $C$ of oc-circuit $\CCC$ is a probabilistic circuit, where uniformly random strings, $r_i \uset \{0,1\}^{N_r}$ for $i=1,..,K$, are input to $C$ and the output of $\CCC$ is distributed over the random space of $\{r_i\}_{i=1,..,K}$.
   
Here note that $\{r_i\}_{i=1,..,K}$, which is an input to $C$, is not included in $\CCC$, while the other inputs to $C$, $u$ and $\{m_i\}_{i=1,..,K}$, are included in $\CCC$. 
In other words,
the size of the randomness,
$\sum_{i=1}^K |r_i|$, is ignored in the size of $\CCC$ or the definition of the organized complexity (see Definition \ref{def:oc}), while $N_r$ and a part of $C$ regarding the randomness are included in $\CCC$. This is because
%, in the concept of the organized complexity,
the randomness, $\{r_i\}_{i=1,..,K}$, is just the random source of $\CCC$'s output distribution and has no organized complexity itself. Hence, simply random objects are characterized to have low organized complexity based on the size of oc-circuit $\CCC$ (see item 3 in the property summary of the proposed complexity measure in the end of this section). 
%is freely available without increasing the organized complexity.
%while universe $u$ and semantics $\{m_i\}_{i=1,..,K}$ are key sources of organized complexity.  

\end{remark}
  
\begin{remark}
\label{rmk:naming}
Although the parts of an oc-circuit, $\overline{C}$, $u$, and $\vec{m}$, are named logic, universe, and semantics, respectively,
we do not care about the meanings of these names in Section \ref{sec:proposed}.
We care more about these meanings in Section \ref{sec:proposed-sit}.

\end{remark}

\begin{definition}(Organized Complexity)
\label{def:oc}

Let $X$ be a distribution over $\{0,1\}^n$ 
for some $n \in \N$.

``Organized complexity'' $\OC$ of distribution $X$ 
at precision level $\delta$ ($0\leq \delta < 1$) is 
\begin{eqnarray}
\label{eq:oc}
&&
\OC(X,\delta) := 
\min\{|{\cal C}| \mid 
X \scl 
Y \rset {\cal C}\}, \ 
\end{eqnarray}
where 
${\cal C} := 
(\overline{C}, u, n, \vec{m})$
is an oc-circuit, and 
$|{\cal C}|$ denotes the bit length 
of the binary expression of ${\cal C}$
(see Section \ref{sec:notation} for the notations of $\scl$). %and $(....)_n$).

We call oc-circuit $\CCC^X \set$ 
$(\overline{C}^X, u^X, n, \vec{m}^X)$
the shortest (or proper) oc-circuit of $X$ at precision level $\delta$,
if
$X \scl Y \rset {\cal C}^X$ and
$|{\cal C}^X| = \OC(X,\delta)$.
If there are multiple shortest oc-circuits of $X$, i.e., they have the same bit length,
the lexicographically first shortest one
is selected as the shortest oc-circuit. 

Then, $\overline{C}^X$ 
$u^X$, and $\vec{m}^X$ 
are called the ``proper logic,'' ``proper universe,'' and
``proper semantics'' of $X$ at precision level $\delta$, respectively.
Here, $X \scl Y \rset \CCC^X \set 
(\overline{C}^X, u^X, n,\vec{m}^X)$.
  
\end{definition}

\begin{theorem}
\label{thm:computable}
For any distribution $X$ over $\{0,1\}^n$ ($n \in \mathbb{N}$)
and any precision level $\delta > 0$,
$\OC(X,\delta)$ can be computed.
\end{theorem}

\noindent
{\bf Proof}

For any distribution $X := \{ (x, p_x) \mid x \in \{0,1\}^n, %\ 0\leq p_x \leq 1, 
\ p_x \in [0,1], \ 
\sum_{x \in \{0,1\}^n} p_x = 1 \}$
and any precision level $\delta >0$,
there always exists another distribution 
$X' := \{ (x, p'_x) \mid x \in \{0,1\}^n, 
\ 0\leq p'_x \leq 1, \ p'_x \in \Q, \ 
\sum_{x \in \{0,1\}^n} p'_x = 1 \}$
such that
$X' \overset{\delta}{\approx} X$
(Here note that $p_x \in \R$ is changed to  
$p'_x \in \Q$
provided that  
$X' \overset{\delta}{\approx} X$). 

We then construct the truth table of Boolean function 
$f: \{0,1\}^\ell \rightarrow \{0,1\}^n$ such that 
$\Pr[x = f(r) \mid r \uset \{0,1\}^\ell] = p'_x$
for all $x \in \{0,1\}^n$, where
the probability is taken over $\uset \{0,1\}^\ell$.
Such a function, $f$, can be achieved by setting truth table 
$T_f :=  \{(r, f(r))\}_{r \in \{0,1\}^\ell}$
such that ${\#}\{ r \mid f(r) = x \} / 2^\ell = p'_x \in \Q$
for all $x \in \{0,1\}^n$.

Since any Boolean function can be achieved by a circuit with basis 
$B$ := \{AND, OR, NOT\} \cite{Enderton01},
we construct circuit $C^*$ for oc-circuit ${\cal C}^*$ 
with
$N_s:=1, N_u=N_m:=0$ (i.e., $u=m_i:=\lambda$), $N_r := \ell$,
$K:=1$, $L_y:= n$, $s_1=s_2:=0$.
That is,
$(0,y_1) := C^*(\lambda, 0,\lambda, r_1)$,
and the output of ${\cal C}^*$ is $y_1 \in \{0,1\}^n$
with the same distribution as that of $X'$
over the randomness of $r_1 \uset \{0,1\}^{N_r}$.
That is, $y_1 \rset {\cal C}^*$ and 
$X \overset{\delta}{\approx} X' = y_1$. 

From the definition of $\OC$,
$\OC(X,\delta) \leq |{\cal C}^*|$.

We then, exhaustively check all values of $Z$ with $|Z| < |{\cal C}^*|$
whether $Z$ is an oc-circuit such that 
$
X \overset{\delta}{\approx} 
Y \rset Z.
$   
Here note that we can syntactically check  
whether or not $Z$ is the correct form of an oc-circuit.
Finally, we find the shortest one among the collection
of $Z$ (and ${\cal C}^*$) satisfying the condition.
Clearly, the size of the shortest one is $\OC(X,\delta)$. 
\qed
\vspace{7pt}

\begin{remark}
\label{rmk:oc-circuit}
As clarified in this proof,
given object (distribution) $X$ and precision level $\delta$,
the proposed definition of organized complexity
uniquely determines (computes)
not only organized complexity $\OC(X,\delta)$
but also the shortest (proper) oc-circuit, $\CCC^X$, including
proper logic $\overline{C}^X$, proper universe $u^X$, and
proper semantics $\vec{m}^X$ of $X$. 
In other words,
the definition
characterizes the complexity features of object $X$,
i.e., it characterizes not only organized complexity $\OC(X,\delta)$,
but also structural complexity features
of $X$,
e.g., computational and distributional features 
by the size of $\overline{C}^X$ and 
descriptional features by the size of $u^X$ and $\vec{m}^X$.
\end{remark}

In the following theorem,
we show that the notion of oc-circuit with the proposed organized
complexity includes the $\epsilon$-machine with 
statistical complexity introduced in Section \ref{sec:survey} 
as a special case.

\begin{theorem}
\label{thm:sim-epsilon}
Any $\epsilon$-machine can be simulated by an oc-circuit.
\end{theorem}

\noindent
{\bf Proof}

Given $\epsilon$-machine,
$(\{S_1, ..., S_k\},
(T_{ij})_{i=1,..,k; j=1,..,k},$
$(\sigma_{ij})_{i=1,..,k; j=1,..,k}
)$,
we construct oc-circuit ${\cal C}$ $:=$ 
$(({C}, N_u, N_s, N_m, N_r, L_y, s_1), u, n,
(m_1,m_2,...))$
such that
$N_s :=$ $\lceil \log_2{k} \rceil$ $+1$
(i.e., $S_i \in \{0,1\}^{N_s})$, 
$N_u := 0$,
$N_m := 0$,
$N_r := \max_{i,j}\{ |T_{ij}| \},$
$L_y := \max_{ij}\{ |\sigma_{ij}| \},$
$s_1 := S_1$ (initial causal state),
$n := \infty$,
$u :=$ $\lambda$ (null string), 
$m_\iota :=$ $\lambda$ ($\iota=1,2,...$) and
$C$ is achieved to satisfy
$\Pr[(S_j,\sigma_{ij}) := C(\lambda, S_i, \lambda, r)] = T_{ij}
$
for $i,j = 1,..,k$,
where 
$|T_{ij}|$ is the bit length of the binary expression of $T_{ij}$, and 
the probability is taken over the 
randomness of $r \uset \{0,1\}^{N_r}$
in each execution of $C$.
It is clear that the behavior of this oc-circuit
with respect to the causal states 
is exactly the same as that for the given $\epsilon$-machine.
\qed

\begin{remark} (Features of the proposed complexity) \ \
\label{rmk:statistical complexity}
The proposed organized complexity is characterized by 
the minimum length of the description of whole oc-circuit ${\cal C}$,
but the statistical complexity is characterized by some partial information on ${\cal C}$, i.e., only the average size of a compressed 
coding of a causal state, $H(D_S) \leq N_s$, 
for $\alpha=1 \ (C_1)$, 
or the (uncompressed) size of a causal state, $N_s$, for $\alpha=0 \ (C_0)$.
That is, our complexity measure captures the complexity of whole circuit (logic) $\overline{C}$, while the statistical complexity
only captures a partial property of the
``distributional complexity'' of $\overline{C}$,
the compressed or uncompressed size of a causal state,
but ignores the distributional complexity 
of $T_{ij}$ and $\sigma_{ij}$ 
(expressed by $N_r$ and the complexity of $\overline{C}$).
That is, even if we only focus on  
the distributional complexity features
(where $N_u=N_m=0$),
the statistical complexity only captures some of the features,
while our definition captures
the whole as the size 
of $\overline{C}$ including $N_s$
and $N_r$.
  
In addition, our complexity measure can treat more general cases with 
$N_u > 0$ and $N_m > 0$,
while statistical complexity only considers 
a limited case with 
$N_u = 0$ and $N_m = 0$, i.e.,
it ignores the descriptional complexity features as well as 
the computational features.
For example, a sequence in a genome pattern that is 
common to all individuals is considered to be determined 
in the evolution process, and has some biological meaning.
The biological knowledge of DNA necessary to 
understand the DNA sequences can be captured by 
logic $\overline{C}$ and universe $u$ of
the oc-circuit $\CCC$ (where $|u| = N_u > 0$), and
the characteristic information (biological meaning) 
on each genome pattern ($\scl Y \rset \CCC)$ 
can be captured by semantics $\vec{m}$ 
of $\CCC$
(where $|m_i| = N_m > 0$).  

Moreover, our complexity definition covers more complexity features. 
The ``computational complexity'' features of an object, which are captured by the logical depth, are characterized by
the size of logic $\overline{C}$
of oc-circuit $\CCC$ in our definition, 
and
the ``descriptional complexity'' features 
of an object, which are captured by the effective complexity (and Kolmogorov complexity), are characterized by the size of universe $u$ and
semantics $\vec{m}_n$ 
of the oc-circuit $\CCC$.
\end{remark}

We can achieve a circuit $C$ using another basis of gates, 
e.g., \{NAND\} and \{AND, NOT\}, 
in place of \{AND, OR, NOT\}.
We express an oc-circuit using such a basis
by 
${\cal C}_{\{\rm NAND\}}$ and ${\cal C}_{\{\rm AND,NOT\}}$, 
respectively.   
We also express the organized complexity of $X$ using such a circuit
by $\OC_{\{\rm NAND\}}(X,\delta)$ and $\OC_{\{\rm AND,NOT\}}(X,\delta)$, 
respectively.

Based on such a different basis of gates,  
a natural variant of the proposed organized complexity,
{\it structured organized complexity},
is given below.

\begin{definition}(Structured Organized Complexity)
\label{def:soc}

Let $X$ be a distribution over $\{0,1\}^n$ 
for some $n \in \mathbb{N}$.

Let
${\cal C}^{\sf S} := 
(\overline{C}^{\sf S}, u, n, \vec{m})$
be a structured oc-circuit,
where
$\overline{C}^{\sf S}$
$= 
({\sf macros}, 
{C}_{{\sf macros}}, 
N_u, N_s, N_m, N_r, L_y, s_1)$
and 
${\sf macros}$ represents a set of macro gates (subroutine circuits) that 
are constructed from basis gates and that can be hierarchically constructed, where
a level of macro gates are constructed from
lower levels of macro gates. 
In addition, ${\sf macros}$ is notationally abused as the canonical 
description of macro gates in ${\sf macros}$,
which is specified in the same manner as 
that in the canonical description of a circuit.
Term ${C}_{{\sf macros}}$ is (the canonical description 
of) a circuit
constructed from basis gates $B$ as well as macro gates in ${\sf macros}$.

Structured organized complexity $\OC^{\sf S}$ of distribution $X$ 
at precision level $\delta$ ($0\leq \delta < 1$) is 
\begin{eqnarray*}
&&
\OC^{\sf S}(X,\delta) := 
\min\{|{\cal C}^{\sf S}| \mid 
X \overset{\delta}{\approx}
Y\rset {\cal C}^{\sf S}\}.
\end{eqnarray*}

\end{definition}

\begin{theorem}
\label{thm:computable-str}
For any distribution $X$ over $\{0,1\}^n$ ($n \in \mathbb{N}$) and any precision level $\delta > 0$,
$\OC^{\sf S}(X,\delta)$ can be computed.
\end{theorem}

\noindent
{\bf Proof}
Given distribution $X$, we can construct  structured oc-circuit ${\cal C}^*$
in the same manner as that shown in the proof of Theorem
\ref{thm:computable}.
Here note that any (basic) oc-circuit ${\cal C}$
can be expressed as structured oc-circuit ${\cal C}^{\sf S}$
where ${\sf macros}:= \lambda$, with slightly relaxing the format for structured oc-circuits, or to allow ${\sf macros}:= \lambda$.

From the definition of $\OC^{\sf S}$,
for any value of $0 < \delta < 1$,
$\OC^{\sf S}(X,\delta) \leq |{\cal C}^*|$.

We then, given $\delta$, exhaustively check all values of $Z$ with $|Z| < |{\cal C}^*|$
whether $Z$ is a structured oc-circuit such that 
$X \overset{\delta}{\approx}
Y \rset Z$.
Finally, we select the shortest one among the collection
of $Z$ (and ${\cal C}^*$) satisfying the condition.
Clearly, the size of the shortest one is $\OC^{\sf S}(X,\delta)$. 
\qed

\begin{remark}
\label{rmk:s-oc-c}
As described in Remark \ref{rmk:oc-circuit},
the shortest structured oc-circuit with these parameters
characterizes the properties of obeject $X$.
It especially shows the optimized hierarchically structured circuit
${\cal C}^{\sf S}$.
\end{remark}

\begin{remark}
\label{rmk:s-oc-example}
As mentioned above, 
we can construct a structured oc-circuit using another basis such as
\{NAND\} and \{AND, NOT\}, e.g., 
${\cal C}^{\sf S}_{\{\rm NAND\}}$ and ${\cal C}^{\sf S}_{\{\rm AND,NOT\}}$.
Then, ${\sf macros}$ in ${\cal C}^{\sf S}_{\{\rm NAND\}}$ can consist of 
macro gates of AND, OR, and NOT from NAND gates.
Since the size of ${\sf macros}$ is a constant $O(1)$ in $n$,
$$
\OC^{\sf S}(X,\delta) \leq \OC^{\sf S}_{\{\rm NAND\}}(X,\delta) \leq 
\OC^{\sf S}(X,\delta) + O(1).
$$

\end{remark}

\vspace{5pt}
\noindent
{\bf Variations:}
\label{phrase:variations}

We have more variations of the organized complexity.

\begin{enumerate}
\item (Computational distance)
In Definitions \ref{def:oc} and \ref{def:soc},
statistical distance is used for defining the closeness
$\overset{\delta}{\approx}$.
We can replace this with the ``computational'' closeness
$\overset{(\DDD,\delta)}{\approx}$.
Here, for two distributions, $X$ and $Y$, over $\{0,1\}^n$, 
the computational closeness of $X$ and $Y$
is defined by \\
$X \overset{(\DDD,\delta)}{\approx} Y$  \ \ iff \ \  
$
\forall D \in \DDD,
$ \ \ 
$ 
\frac{1}{2}\cdot | \Pr[1 \rset D(\alpha) \mid \alpha \rset X]
- \Pr[1 \rset D(\alpha) \mid \alpha \rset Y] |
< \delta,
$

where $\DDD$ is a class of machines $D: \{0,1\}^n \ra \{0,1\}$. 
Intuitively, the computational closeness means that $X$ and $Y$ are
indistinguishable at precision level $\delta$ by any machine in class $\DDD$.

\item (Quantum circuits)
Circuit $C$
in Definitions \ref{def:oc} and \ref{def:soc}
can be replaced by a ``quantum'' circuit \cite{NieChu00}.
In this variation, we assume that the source of a distribution (observed data) is 
principally given by a quantum phenomenon. 

There are several variations of the quantum complexity definition, typically:
(1) all inputs and outputs of $C$ are classical strings,
(2) only state $s_i$ is a quantum string and the others are classical, and
(3) all inputs and outputs of $C$ except output $y_i$ are quantum strings.

\item (Interactions)
If an observation object actively reacts similar to a living thing, we often observe it in an interactive manner.

So far in this paper we have assumed that an object is a distribution that we perceive passively. 
We can extend the object from such a passive one to an active one with interactive observation.

Suppose that the observation process is interactive between observer $A$ and observation object $B$. For example, 
$A$ first sends $z_1$ to $B$, which replies $x_1$ to $A$, and we continue the interactive process, $z_2$, $x_2$, ..., $z_J$, $x_J$.

Let $Z \set (z_1,..,z_J)$ and $X \set (x_1,.., x_J)$ be distributions.  
We then define the conditional organized complexity of $X$ under $Z$ with precision level $\delta$, ${\sf OC}(X : Z, \delta)$,
which can be defined as the shortest (finite version of) conditional oc-circuit to simulate $X$ (with precision level $\delta$) under $Z$   
(see Definition \ref{def:cond-occ}
for the conditional oc-circuit).

The extended notion of the organized complexity of interactive object $(X,Z)$ can be defined by ${\sf OC}(X : Z, \delta)$. 
     
\end{enumerate}

\vspace{5pt}

In light of the criteria described in Section \ref{sec:criteria},
the proposed complexity measure has the following properties.

\begin{enumerate}
\item
The proposed complexity definition covers probability
distributions and deterministic strings (as special cases of 
distributions) in a unified manner (Good).
 
For example, for any deterministic string (as a special case of distribution), there exists an oc-circuit with circuit $C$ to simulate the deterministic string by
$Y \set (y_1,..,y_K)_n \in \{0,1\}^n$, 
%is expressed by an oc-circuit with circuit $C$ 
such that
\begin{eqnarray*}
&&
(s_{i+1}, y_{i}) := C(u, s_{i}, m_i, \lambda), \ \ i=1,2,..., K, 
\end{eqnarray*}
where $N_r = 0$, i.e., $r_i := \lambda$ for $i=1,..,K$.

\item
Simple (or very regular) objects have low complexity (Good).

For example, in a very regular case (`$11\cdots1$' $\in \{0,1\}^n$),
the object can be simulated by an oc-circuit $\CCC \set 
(\overline{C}, 1, n,\lambda)$
with $\overline{C} \set (C, 1, 1, 0, 0, 1, 0),$
such that
$$
(s_{i+1}, 1) := C(1, s_{i}, \lambda, \lambda), \ \ i=1,2,..., n, 
$$
where $N_u=1 (u=1)$, $s_i:=0$ for $i=1,..,n+1$ (i.e., $N_s=1$), 
$N_m = N_r = 0$ (i.e., $m_i = r_i = \lambda$) and $L_y=1$.
That is, input size $N = 2$ and output size $L=2$, i.e.,
$C$ has 2 gates that are input gates labeled by $(1,2)$
and that are also output gates. 
${C} :=  
((w_{ij}) := I_2, 
(\ell_1,\ell_2):=(1,2), (o_1,o_2):=(2,1))$,
where $I_2$ is the 2-dimensional identity matrix.
Hence, 
${\cal C} := (( ((1,0,0,1),(1,2),(2,1)),
 1, 1, 0, 0, 1, 0 ), 1, n,\lambda)$, and
$\OC($`$11...1$'$) \leq c + \log_2{n}$,
where $c$ is a small constant.

\item
Simply random objects have low complexity (Good).

For example, in the case of uniformly random distribution $X$
over $\{0,1\}^n$, 
the object can be simulated by an oc circuit $\CCC \set 
(\overline{C}, \lambda, n,\lambda)$
with $\overline{C} \set (C, 0, 1, 0, 1, 1, 0),$
such that  
$$
(s_{i+1}, r_i) := C(\lambda, s_{i}, \lambda,r_i), \ \ i=1,2,..., n, 
$$
where $N_u=0$, $s_i:=0$ for $i=1,..,n$ (i.e., $N_s=1$), 
$N_m=0$ (i.e., $u=m_i=\lambda$), $N_r = 1$,
$r_i \uset \{0,1\}$, and $L_y=1$.
That is, input size $N = 2$ and output size $L=2$, i.e.,
$C$ has 2 gates that are input gates labeled by $(1,2)$
and that are also output gates. 
${C} :=  
((w_{ij}) := I_2, 
(\ell_1,\ell_2):=(1,2), (o_1,o_2):=(1,2))$.
Hence, 
${\cal C} := ((((1,0,0,1),(1,2),(1,2)), 
0, 1, 0, 1, 1, 0), \lambda, n, \lambda)$, and
$\OC(X) \leq c + \log_2{n}$,
where $c$ is a small constant. 

\item 
Highly organized objects have high complexity (Good).

As described in Remarks \ref{rmk:oc-circuit} and \ref{rmk:statistical complexity},
the proposed complexity definition simultaneously 
captures the distributional features of  complexity 
(similar to statistical complexity and effective measure complexity), 
computational features of complexity (similar to logical depth), 
and descriptional features of complexity
(similar to the Kolmogorov complexity and effective complexity).

Hence, our complexity measure does not have the drawbacks of the existing complexity
measures described in Section \ref{sec:survey},
i.e., our measure correctly evaluates the complexity of highly 
organized objects for which the existing measures 
miss-evaluate to be low.
In addition, objects evaluated by any existing 
(organized) complexity measure to be high for some features of complexity are also evaluated to be high 
in our complexity measure.

\item
The proposed complexity of any object (distributions/strings) 
is computable as shown in Theorems \ref{thm:computable} and
\ref{thm:computable-str} (Good).

\end{enumerate}

%%%%% 20160703
  
\section{Semantic Information Theory}

\subsection{Existing Theories and Problems}  
\label{sec:existing-semantic-theories}

As described in Section \ref{sec:bacgroud},
the semantic information theory has been 
studied for over six decades, 
e.g., \cite{BaoBasDeaParSwaLel11,Bacchus88,CarBar52,D'alfonso11,Flordi04,Flordi09,
JubSud08i,JubSud08ii,KohSch09,Langel09,NafAle09,Nillsson86,Sommagura09,
WilKal05}.
Among these studies, we here investigate the existing semantic information theories 
that offer a quantitative measure of semantic information,
e.g., \cite{BaoBasDeaParSwaLel11,Bacchus88,CarBar52,D'alfonso11,Flordi04,Flordi09,
NafAle09,Nillsson86,Sommagura09}.

Although they present many different ideas and approaches,
a common paradigm among these theories 
assumes some a priori information, 
e.g., world model, knowledge database, and logic,
to measure the amount of semantic information of an object. 
 
For example, in \cite{BaoBasDeaParSwaLel11},
the following information is assumed to be established beforehand to define the semantic information measure.
 
\noindent
- A world model that is a set of interpretations for
propositional logic sentences with probability distributions.
 
\noindent
- An inference procedure for propositional logic.

\noindent
- A message generator that generates messages using some coding strategy.

It looks natural and inevitable to assume such a priori basic information 
such as the world model and logic to
define the semantic information measure, but
we have the same criticism for this paradigm as that 
for the thermodynamic depth \cite{LloPag88} 
described in Section \ref{sec:survey}.
That is, any existing semantic information theory 
in literature assumes such a priori information 
but gives no concrete or precise specification 
of the assumed a priori information. 
Without any concrete specification of
the assumed information, we cannot rigorously define the semantic
information amount 
and such a quantitative definition is just a vague and obscure notion.
If there are thousands of possible concrete specifications of the information, 
we would have thousands of possible quantitative definitions.

In addition, 
when we try to measure the semantic information amount of an object, or we have no idea of its amount,
such a priori information should be established  beforehand and its complexity (information amount) should be comparable to or exceed  
that of the object.
Hence, if the a priori information is fixed, or
the semantic information measure with this information is concretely defined, it cannot measure the information amount of an object that has greater information amount than that of the a priori information.  
That is, any concrete definition in this paradigm can measure only a specific subset of objects, i.e., any concrete and generic definition is impossible, or any concrete definition is ad hoc.     
It should be an essential problem in the existing semantic information theories.

Another criticism of the existing semantic information theories
is that they are constructed only on some mathematical logic such as propositional   
and first order logics. 
Our daily communications should be based on more complicated and fuzzy logic.
It is well known that bees inform other bees of the direction and distance of flowers 
using their actions similar to dancing. Clearly some semantic information is transferred
among bees in this case, and the logic for the semantics should be 
much different from mathematical logic and the logic of humans.

Given an object, e.g., Shakespeare's plays
and bee's actions,
we may roughly imagine which 
classes of information (universe) and logic are necessary 
or sufficient to understand the object.
For example, in order to understand the semantic information 
of Shakespeare's plays, the necessary 
universe and logic 
may be the knowledge of English sentences, the culture of that age and human daily logic.
To understand the semantic information of bee communications,
a much more limited and specific type of 
universe and logic 
may be sufficient.

We now have the following questions.
\begin{enumerate} 
\item
Can we quantitatively define the amount of semantic information  
of an object without assuming any a priori information?
Or, can we quantitatively define the amount of semantic information  
of an object absolutely (not relative to
a priori information)? 

\item
Given an object, can we determine the minimum 
amount of universe and logic
to understand the semantic information of the object?  

\end{enumerate}

\subsection{Proposed Semantic Information Theory}  
\label{sec:proposed-sit}

In this section, we present a mathematical theory of {\it semantic}
information and communication
that covers the semantic and effectiveness problems
(Levels B and C of information and communication problems given by
Weaver, see Section \ref{sec:bacgroud}). 
The proposed theory is 
based on our organized complexity measure shown in Section
\ref{sec:proposed}.
The proposed semantic information theory offers 
a positive answer to the questions 
raised at the end of Section \ref{sec:existing-semantic-theories}.

We first consider an example described in Section \ref{sec:criteria},
a source (distribution) from
Shakespeare's plays.

Let distribution $X$ over $\{0,1\}^n$ be the source of 
several plays
that consists of several hundred pages of English sentences, and
let ${\cal C}^X$ be the shortest (proper) oc-circuit, 
$(\overline{C}^X, u^X, n, \vec{m}^X)$
to simulate source $X$ at precision level $\delta$.
The output, $Y \rset (\overline{C}^X, u^X, n,\vec{m}^X)$,
is the distribution statistically $\delta$-close to $X$,
the distribution of the source 
of the whole sentences in the plays.

The shortest oc-circuit, ${\cal C}^X$, for $X$ can characterize source $X$ such
that
\begin{itemize}
\item
the proper logic, 
$\overline{C}^X$ 
of $X$
may capture the features of Shakespeare's way of thinking
and daily logic including English grammar,
\item
the proper universe of $X$, $u^X$,  
may capture the knowledge of English words 
and expressions as well as aspects of 
the cultures necessary to understand the plays,
\item
the proper semantics of $X$, $\vec{m}^X$, 
may capture the semantics (meanings) of sentences of the plays.
\end{itemize}

Based on the observation above, we formalize 
the semantic information theory.
 
In Section \ref{sec:proposed}, we aim to define quantitatively   
the organized complexity of {\it physical} objects,
i.e., the sources of physically observed data.
Since any physical thing is essentially bounded finitely,
we assume a source is a distribution over finite strings,
$\{0,1\}^n$ ($n \in \N$),
in Section \ref{sec:proposed}.

In contrast, in this section, we aim to construct 
a {\it mathematical} theory of semantic information,
where we focus on the {\it asymptotic} properties of an object 
when the size of the object is supposed to be increasing unlimitedly.
This is because this theory focuses on the semantics part of the organized complexity, which consists of logic, universe, and semantics.  
The semantics part such as 
the proper semantics, $\vec{m}^X$, of object (source) $X$ in the example above
can be characterized   
more clearly and simply using the asymptotic properties (where the sizes of $\vec{m}^X$ and $X$ are supposed to be increasing unlimitedly while the sizes of proper logic $\overline{C}^X$ and proper universe $u^X$ are finitely bounded)   
than by the finite-size properties (where the size of $\vec{m}^X$ is finitely bounded).
Note that Shannon's information theory is 
also a theory for asymptotic properties.
Therefore, an object here is not a single 
distribution but a {\it family} of distributions,
$\XXX \set \{ X_n \}_{n \in \N}$,
where $X_n$ is a distribution over $\{0,1\}^n$.

We define several notions 
including the {\it semantic information amount}.
Note that Occam's razor also plays a key role in this definition,
since it is based on organized complexity,
${\sf OC}$.  

\subsubsection{Semantic Information Amount}

First we introduce the notion of a {\it family of distributions}
and (naturally) extend the concept of an oc-circuit 
(for a distribution on finite-size strings) 
into that of an {\it oc-circuit for 
a family of distributions},
which is the same as 
the original one 
except that
its output and semantics are unbounded in this concept,
while they are bounded in the original.

\begin{definition} (Family of Distributions and 
OC-Circuit for a Family of Distributions) 
\label{def:dist-family} 

Let $X_n$ be a distribution such that 
$X_n := \{ (x, p_x) \mid x \in \{0,1\}^n, \ 0\leq p_x \leq 1, \ 
\sum_{x \in \{0,1\}^n} p_x = 1 \}$,
and 
${\cal X} \set \{X_n\}_{n \in \mathbb{N}}$ be a ``family of distributions.''

Let ${\CCC} := 
(\overline{C}, u, \infty,  \vec{m}_\infty)$ 
be an ``oc-circuit for a family of distributions'' 
such that
\begin{eqnarray*}
&&
\overline{C} \set (C, N_u, N_s, N_m, N_r, L_y, s_1),
\ \ \ 
\\
&&
\vec{m}_\infty \set 
(m_i)_{i=1,2,...} \set 
(m_1,m_2,...) \in \{0,1\}^\infty, \ \ 
\\
&&
(s_{i+1}, y_{i}) \longleftarrow 
\fbox{$C(u, \ \cdot \ )$}
\longleftarrow (s_{i}, m_i, r_i), \ \ i=1,2,..., 
\\
&&
\mbox{i.e.,} \ 
(s_{i+1}, y_{i}) := C(u, s_{i}, m_i, r_i),  \ \ i=1,2,...,
\\
&&
Y_n  \set 
(y_1,..,y_{K_n})_n, 
\ \ \  
K_n \set \lceil n/L_y \rceil, 
\ \ \mbox{for} \ n \in \N,
\\
&&
\YYY \set \{Y_n\}_{n\in\N} \rset 
\CCC, 
\end{eqnarray*}
where  
$N_m \leq L_y$,
$m_i \in \{0,1\}^{N_m}$,
$u \in \{0,1\}^{N_u}$,
$s_i \in \{0,1\}^{N_s}$,
$r_i \in \{0,1\}^{N_r}$,
$y_i \in \{0,1\}^{L_y}$
and
$\{0,1\}^\infty$ is the set of infinite-size binary strings. 

Let ${\CCC}_n := 
(\overline{C}, u, n, \vec{m}_n)$,
where
$\vec{m}_n \set (m_1,..,m_{K_n})$.

\end{definition}

\begin{definition}
\label{def:seq-fd}
(Sequential Family of Distributions)

A family of distributions, 
$\YYY \set \{Y_n\}_{n\in\N}$,
is called a 
``sequential family of distributions''
if
for any $n'$ and $n$ in $\N$ with $n' > n$,
distribution $Y_n$ over $\{0,1\}^n$
is the $n$-bit restriction
of distribution $Y_{n'}$ over $\{0,1\}^{n'}$
(see Section \ref{sec:notation} for
the definition of $n$-bit restriction).

\end{definition}

\begin{remark}
The family of distributions output by 
an oc-circuit for a family of distributions,
$\YYY \set \{Y_n\}_{n\in\N}$,
is a sequential family of distributions.

A value of 
$\mu \set(\mu_1, \mu_2, ...) 
\in \{0,1\}^\infty$ 
\ ($\mu_i \in \{0,1\}$ for $i=1,2,...$)
corresponds to a value in 
$[0,1] \subset \R$
by map 
$\varphi : \{0,1\}^\infty  \mapsto [0,1]$, \  
$\varphi: (\mu_1, \mu_2, ...) 
\ra ``0.\mu_1 \mu_2 ...$''  $\in [0,1]$,
where $``0.\mu_1 \mu_2 ...$'' is the 
binary expression of a value in
$[0,1]$.

Through this correspondence,
if $\YYY \set \{Y_n\}_{n\in\N}$
is a sequential family of distributions,
$\lim_{n\ra \infty} Y_n$
corresponds to probability density function  $Y(\cdot)$ 
with support $[0,1]$ \cite{CovTho91}
such that
$\displaystyle{\int_0^1 Y(x) dx = 1}$, and 
$Y_n = \{ (x, p_x) \mid
x \in \{0,1\}^n, \ 
p_x \set \displaystyle{
\int_{``0.x{\textrm "}}^{``0.x{\textrm "}+1/2^n} Y(x) dx} \}$,
where 
$``0.x$" denotes $``0.x_1 x_2 ... x_n$"
$\in [0,1]$, if $x = (x_1,..,x_n) \in \{0,1\}^n$.

\end{remark}

Since there are a variety of unnatural or eccentric
distribution families in general,
we introduce a class of distribution families,
{\it semantic information sources}, that are
natural or appropriate as the object of the semantic information theory.   

Although a family of distributions 
covers an unbounded number of distributions,
the core mechanism, e.g., logic and universe, of 
a source to produce a family of distributions 
should be {\it bounded} due to the physical constraints. 
In other words, such a natural family of distributions 
should be actualized as an unbounded series of distributions produced by a physically bounded mechanism, e.g., logic and universe, along with an unbounded series of 
inputs, e.g., semantics.

Since oc-circuits are sufficiently general to express 
any distribution 
(as shown in Theorem \ref{thm:computable}),
a natural and appropriate object in the semantic information theory should be expressed by 
an oc-circuit for a family of distributions, which is defined in
Definition \ref{def:dist-family}. 

We have another condition for an appropriate object or its oc-circuit. 
Roughly, the shortest oc-circuit 
for simulating an appropriate object
should be converging asymptotically,
since we aim to characterize an object 
by the asymptotic properties 
in our theory.
Then,
the shortest expression 
(logic, universe, and semantics)
of an oc-circuit simulating 
a {\it family of distributions}, $\XXX \set \{X_n\}$,
should be equivalent to the shortest one for
{\it each distribution} $X_n$ 
asymptotically (for sufficiently large $n$).     
Namely, the {\it global} 
shortest expression 
(the proper logic, universe, and semantics
of $\XXX$)
should be equivalent to the {\it local} 
shortest expression 
(the proper logic, universe, and semantics
of $X_n$) asymptotically. 
 
\begin{definition} (Semantic Information Source) 
\label{def:sis}

A family of distributions, ${\cal X} \set \{X_n\}_{n \in \mathbb{N}}$, is called a 
``semantic information source''
at precision level $\delta(\cdot)$
if 
there exists an oc-circuit for a family of distributions,
${\CCC}^{\XXX} := 
(\overline{C}^{\XXX}, u^{\XXX}, \infty,  \vec{m}^{\XXX}_\infty)$,
where ${\CCC}^{\XXX}_n := 
(\overline{C}^{\XXX}, u^{\XXX}, n,  \vec{m}^{\XXX}_n)$,
and $\vec{m}^{\XXX}_n$ is the 
($N^{\XXX}_m \cdot \lceil n/L_y^{\XXX}\rceil$-bit)
prefix of $\vec{m}^{\XXX}_\infty$
for $n$-bit output,
that satisfies the following conditions.
\begin{itemize}
\item
For all $n \in \N$, \ \ 
$
X_n \overset{\delta(n)}{\approx}
Y_n^{\XXX} \rset \CCC^{\XXX}_n
$,
and 
\item 
there exists $n_0\in \mathbb{N}$ 
such that
for all $n \geq n_0$,
\begin{eqnarray}
&& 
| (\overline{C}^{\XXX}, u^{\XXX}, \vec{m}^{\XXX}_n) |
=
\min\{ 
|(\overline{C}, u, \vec{m}_n)|
 \ \mid \
\nonumber
\\
&&
\ \ \ \ \ \ \ \ \ \ \ \ 
X_n \overset{\delta(n)}{\approx} 
Y_n \rset 
(\overline{C}, u, n, \vec{m}_n)
 \}.
\label{eq:sis-2}
\end{eqnarray}
\end{itemize}
If there are multiple  
oc-circuits, $\CCC^{\XXX}$, that 
satisfy the above conditions, 
the lexicographically first one
is selected as $\CCC^{\XXX}$ for $\XXX$.  

Then, 
$\CCC^{\XXX}$, 
$\overline{C}^{\XXX}$, $u^{\XXX}$, and
$\vec{m}_\infty^{\XXX}$ are called the 
``proper oc-circuit,'' 
``proper logic,'' ``proper universe,'' and
``proper semantics'' of $\XXX$
at precision level $\delta(\cdot)$, 
respectively.
Here, $\XXX \overset{\delta(\cdot)}{\approx} \YYY \rset {\CCC}^{\XXX} := 
(\overline{C}^{\XXX}, u^{\XXX}, \infty,  \vec{m}^{\XXX}_\infty)$.

For two semantic information sources, $\XXX \set 
\{ X_n \}_{n \in \N}$ and $\YYY \set \{ Y_n \}_{n \in \N}$,
we say $\XXX$ and $\YYY$ are ``semantically equivalent''
at precision level $\delta(\cdot)$
iff they have the same proper oc-circuit 
$\CCC$ at level $\delta(\cdot)$.
We denote this by 
$\XXX \seq \YYY$.

\end{definition}

\begin{remark}
\label{rmk:sis}
From the definition, 
for sufficiently large $n$ 
($\exists n_0 \ \forall n > n_0$), 
\begin{eqnarray*}
&&
|\CCC^{\XXX}_n| = |\CCC^{X_n}|
= \OC(X_n,\delta(n)),
\end{eqnarray*}
where 
$\CCC^{X_n} \set (\overline{C}^{X_n},  u^{X_n}, n, \vec{m}^{X_n})$ 
is the proper oc-circuit
of $X_n$ at precision level $\delta(n)$ 
(see Definition \ref{def:oc} for the proper 
oc-circuit). 

The left term in Eq. 
(\ref{eq:sis-2}) is fixed by $\XXX$,
while the right term varies with each 
$X_n$.  
This definition says that,
in semantic information source
$\XXX \set \{X_n\}_{n\in\N}$,  
$X_n$ 
for all sufficiently large $n$
is uniformly characterized
by a single oc-circuit $\CCC^\XXX$ 
proper for $\XXX$.

\end{remark}

Semantic information source
$\XXX \set \{X_n\}_{n\in\N}$
is characterized by 
its proper oc-circuit 
$(\overline{C}^{\XXX}, u^{\XXX}, \infty,  \vec{m}^{\XXX}_\infty)$, where
$\vec{m}^{\XXX}_\infty \set (m_i)_{i=1,2,...}$
with $m_i \in \{0,1\}^{N_m^{\XXX}}$.
Since $\vec{m}^{\XXX}_\infty$ includes 
an infinite number of strings in $\{0,1\}^{N_m^{\XXX}}$, any value in $\{0,1\}^{N_m^{\XXX}}$ could be 
$m_i$ for some $i \in \N$, or 
the value of $m_i$ for $i \in \N$ could be 
any value in $\{0,1\}^{N_m^{\XXX}}$.  
Hence,
if $(\overline{C}^{\XXX}, u^{\XXX}, \infty,  \vec{m}^{\XXX}_\infty)$
is the proper oc-circuit of semantic information source $\XXX$,
an oc-circuit $(\overline{C}^{\XXX}, u^{\XXX}, \infty,  
\vec{m}_\infty)$ with any other $\vec{m}_\infty$
could be the proper oc-circuit of a 
semantic information source with 
the proper semantics $\vec{m}_\infty$. 

Therefore, a semantic information source
is characterized by its proper logic
$\overline{C} \set 
(C, N_u, N_s, N_m, N_r, L_y, s_1)$ along with universe $u$ 
in the universe space $\mathbb{U}^{\overline{C}} \set 
\{0,1\}^{N_u}$ and
semantics 
$\vec{m}_\infty$ in the semantics space
$\mathbb{M}^{\overline{C}} \set \{ (m_i)_{i=1,2,..} \ \mid \  
m_i \in \{0,1\}^{N_m} \}$.

Namely, 
the $L_y^{\XXX}$ bit output, $y_i$,
should have $N_m^{\XXX}$ bit semantic information,
i.e.,
as $n$ bit output should have $n N_m^{\XXX}/L_y^{\XXX}$ 
(or its rounded-up integer, 
$\lceil n N_m^{\XXX}/L_y^{\XXX} \rceil$)
bit semantic information.

\begin{definition}(Semantic Information Amount and Semantic Information Space)
\label{def:sa}

Let ${\cal X} \set \{X_n\}_{n \in \mathbb{N}}$ be
a semantic information source whose
proper logic at precision level $\delta(\cdot)$
is 
$\overline{C}^{\XXX}
\set 
(C^{\XXX}, N_u^{\XXX}, N_s^{\XXX}, N_m^{\XXX}, N_r^{\XXX}, 
L_y^{\XXX}, s_1^{\XXX})$.

The ``semantic information amount,'' 
${\SA}$, of $X_n \in \XXX$ at precision level $\delta(\cdot)$ is
defined by
$$\SA(X_n,\delta(\cdot)) \set 
\lceil n N_m^{\XXX}/L_y^{\XXX}\rceil.  \ 
$$

Let 
$\mathbb{M}^{\overline{C}} \set 
\{ (m_i)_{i=1,2,..} \ \mid \  
m_i \in \{0,1\}^{N_m} \}$
be the ``semantic information (meaning) space''
of proper logic 
$\overline{C}
\set 
(C, N_u, N_s, N_m, N_r, 
L_y, s_1)$,
and
$\mathbb{M}^{\overline{C}}_n \set \{ (m_i)_{i=1,..,K_n} \ \mid \  m_i \in \{0,1\}^{N_m} \}$
be ''the 
($N_m \cdot \lceil n/L_y^{\XXX}\rceil$-bit)
prefix of $\mathbb{M}^{\overline{C}}$
for an $n$-bit output.''

Let 
$\mathbb{U}^{\overline{C}} \set \{0,1\}^{N_u}$
be the ``universe space''
of proper logic 
$\overline{C}$.

\end{definition}
 
\noindent
{\bf Examples of Semantic Information Sources}
\label{par:example-sis}

Here we show some examples of 
semantic information sources.

\begin{enumerate}

\item
(Example 1)

Let us employ an example of Shakespeare's plays again, and imaginarily suppose that there are an unbounded number of Shakespeare's plays, but that the logic and universe (knowledge) of 
Shakespeare are bounded.

Let $\XXX \set \{X_n\}_{n\in\N}$ be a sequential family of distributions of
Shakespeare's (unbounded number of) plays.  

Given $X_{n^{(1)}}$ with $n^{(1)} \in \N$,
we compute an oc-circuit $\CCC^{(1)}_{n^{(1)}}$
such that 
$\CCC^{(1)}_{n^{(1)}} \set 
(\overline{C^{(1)}}, u^{(1)},n^{(1)},  
\vec{m}^{(1)}_{n^{(1)}})$
is the shortest (proper) oc-circuit to simulate $X_{n^{(1)}}$ at precision level $\delta$,
i.e.,
$
\OC(X_{n^{(1)}},\delta) = |\CCC^{(1)}_{n^{(1)}}|, \ \ 
$

Next, for some $n^{(2)} > n^{(1)}$, we compute  
the shortest (proper)
oc-circuit $\CCC^{(2)}_{n^{(2)}}\set 
(\overline{C^{(2)}}, u^{(2)},n^{(2)},  
\vec{m}^{(2)}_{n^{(2)}})$ to simulate $X_{n^{(2)}}$ at precision level $\delta$.

If for any $n^{(2)} > n^{(1)}$ \ 
$|(\overline{C^{(2)}}, u^{(2)}, (\vec{m}^{(2)}_{n^{(2)}})_{n^{(1)}})| 
= |(\overline{C^{(1)}}, u^{(1)}, \vec{m}^{(1)}_{n^{(1)}})|
$ where $(\vec{m}^{(2)}_{n^{(2)}})_{n^{(1)}} $ is the $n^{(1)}$-prefix of $\vec{m}^{(2)}_{n^{(2)}}$, 
it could imply that 
$\CCC^{(1)}$, $u^{(1)}$ and $\lim_{n^{(2)} \ra \infty} \vec{m}^{(2)}_{n^{(2)}}$ are the proper logic, universe, and semantics of $\XXX$, respectively.

If for some $n^{(2)} > n^{(1)}$ \
$|(\overline{C^{(2)}}, u^{(2)}, (\vec{m}^{(2)}_{n^{(2)}})_{n^{(1)}})| 
\not= |(\overline{C^{(1)}}, u^{(1)}, \vec{m}^{(1)}_{n^{(1)}})|
$,
let $\CCC^*_{n^{(2)}} \set \CCC^{(2)}_{n^{(2)}}$
as a candidate of the proper oc-circuit of $\XXX$ (up to the size of $n^{(2)}$).

We repeat the procedure for $n^{(i)}$ ($i=3,4,...$)
and update $\CCC^*_{n^{(i)}}$.

If $|(\overline{C^{(i+1)}}, u^{(i+1)}, (\vec{m}^{(i+1)}_{n^{(i+1)}})_{n^{(i)}})| 
\not= |(\overline{C^{(i)}}, u^{(i)}, \vec{m}^{(i)}_{n^{(i)}})|
$
for some $i \in \N$,
it should hold that 
$|(\overline{C^{(i*1)}},u^{(i+1)})|
> 
|(\overline{C^{(i)}},u^{(i)}) |$,
since 
the required logic and universe (knowledge)
to understand the plays should increase
as the amount of plays increases. 

In the updating process of $\CCC^*_{n^{(i)}}$, 
the $i$-th semantics part, $\vec{m}^{(i)}_{n^{(i)}}$, with $X_{n^{(i)}}$ has 
some 
redundancy in light of a longer (more global) context with $X_{n^{(i+1)}}$ 
and such 
redundancy 
could be eliminated 
in $(\vec{m}^{(i+1)}_{n^{(i+1)}})_{n^{(i)}}$
and 
absorbed into
the $(i+1)$-th logic and universe, $(\overline{C}^{(i+1)},u^{(i+1)}))$ (for a longer context), i.e., the logic and universe part should increase in the process, while
the semantics part becomes more compressed (shorter) and closer to 
a uniform one. IN addition, 
block size $N_y^{(i)}$ of the output becomes longer,
where a longer block with a longer context is processed using more complicated logic and a larger universe.

The logic and universe part, $(\overline{C}^{(i)},u^{(i)}))$, of 
oc-circuit ${\CCC}^{(i)}$ 
should be finitely bounded
for any $i \in \N$.
Actually, 
$$|(\overline{C}^{(1)},u^{(1)})| \leq 
|(\overline{C}^{(2)},u^{(2)})| \leq ... \leq 
|(\overline{C}^{*},u^{*})|, 
$$
where $(\overline{C}^{*},u^{*})$ should be the proper logic and universe of $\XXX$.

Hence, there exists $i^*\in\N$
such that 
for any $i > i^*$  
($n^{(i)} > n^{(i^*)})$ 
$|(\overline{C}^{(i)}, u^{(i)}, (\vec{m}^{(i)}_{n^{(i)}})_{n^{(i^*)}})|
= |(\overline{C}^{(i^*)}, u^{(i^*)}, \vec{m}^{(i^*)}_{n^{(i^*)}})|
$, i.e.,
there exists $i^*\in\N$ and
$(\overline{C}^{\XXX}, u^{\XXX},
\vec{m}^{\XXX}_\infty)$
$\set 
(\overline{C}^{(i^*)}, u^{(i^*)}, 
\lim_{n^{(i^*)} \ra \infty} \vec{m}^{(i^*)}_{n^{(i^*)}})$
such that 
for any $i > i^*$ 
($n^{(i)} > n^{(i^*)})$ \ 
$|(\overline{C^{\XXX}}, u^{\XXX}, (\vec{m}^{\XXX}_\infty)_{n^{(i)}})| 
= |
(\overline{C}^{(i)}, u^{(i)}, (\vec{m}^{(i)}_{n^{(i)}})|$.

Thus, 
$\XXX$ is 
a semantic information source and
$\CCC^{\XXX} \set (\overline{C}^{\XXX}, u^{\XXX},\infty, \vec{m}^{\XXX}_\infty)$ is the proper oc-circuit of $\XXX$.

\item
(Example 2)

Let $\CCC$ be an oc-circuit that 
outputs a sequential family of distributions,  
$\XXX \set \{X_n\}_{n\in\N} \rset \CCC$,
such that
$\CCC := (\overline{C}, u, \infty, \vec{m}_\infty)$
and
$\vec{m}_\infty \set \{\vec{m}_n\}_{n\in\N}$,
$\vec{m}_n \uset \{0,1\}^{\lceil nN_m/L_y\rceil}$.

For any precision level $\delta$, 
for sufficiently large $n^* \in \N$,
we can compute $\CCC^*_{n^*} := 
(\overline{C}^*, u^*, n^*, \vec{m}^*_{n^*})$
which is the shortest (proper) oc-circuit to simulate 
$X_{n^*}$ at precision level $\delta$.

Then, there exists $\vec{m}^*_\infty$
with high probability 
such that
$\CCC^* \set (\overline{C}^*, u^*,\infty, \vec{m}^*_\infty)$,
and
for any $n > n^*$ \
$\CCC^*_n \set (\overline{C}^*, u^*,\infty, (\vec{m}^*_\infty)_n)$ is the shortest oc-circuit of $X_n$
at precision level $\delta$,
i.e.,
$\CCC^*$
is the proper oc-circuit of $\XXX$
at precision level $\delta$.

This is because 
$\vec{m}_n \uset \{0,1\}^{\lceil nN_m/L_y\rceil}$
and no more data compression on $\vec{m}_n$
is possible for any sufficiently large 
$n > n^*$
with high probability.

That is, $\XXX$ is   
a semantic information source
with high probability.

The difference between this example and the first example
is that
the unbounded semantics sequence, 
$\vec{m}_\infty$,
in this example is uniformly selected 
from the beginning,
while, in the first example,
the semantics sequence is gradually compressed as size $n$ of distribution $X_n$ becomes longer
in the process of updating $\CCC^*_n$.

\item
(Other Examples)

The information sources modeled in the previous semantic information theories  
in literature (in Section \ref{sec:existing-semantic-theories})
are considered to be ``semantic information sources.''

For example in \cite{BaoBasDeaParSwaLel11},
Fig. 2 shows a model of semantic information communication.
Here, $I_S$ (Inference Procedure) and the syntax and logic part of $M_S$ (Message generator)
can be considered as {\it circuit} $C$ of the oc-circuit and $W_S$ (world model), 
$K_S$ (Background Knowledge) can be considered as {\it universe} $u$ of the oc-circuit,
and the semantics of $M_S$ (Message generator) can be considered as semantics $\vec{m}$
of the oc-circuit.   
That is, the messages from Sender $S$ in Fig. 2 can be modeled as
a source generated by an oc-circuit, or a semantic information source. 

\end{enumerate}

We then consider the following problem.
Given semantic information source 
$\XXX$ 
with $\delta(\cdot)$,
can we compute its proper oc-circuit and the related information?
The answer is no, since $\XXX$ 
consists of 
an infinite number of distributions 
and it cannot be described finitely.

The following theorem however, shows that,
given $X_n \in \XXX$ and $\delta(\cdot)$ 
with a sufficiently large $n$,
we can compute the proper oc-circuit
of $\XXX$.

\begin{theorem}
\label{thm:comp-sis}

For any semantic information source $\XXX$
at precision level $\delta(\cdot)>0$,
given 
$X_n \in \XXX$ and $\delta(n)$ for some $n > n_0$, where $n_0$ is given in Definition \ref{def:sis},
the proper oc-circuit
(proper logic, proper universe, and
$n$-prefix of proper semantics) 
of $\XXX$ at precision level $\delta(\cdot)$ 
can be computed.

\end{theorem}

The proof of this theorem
is essentially the same as that for Theorem \ref{thm:computable}.

\begin{remark}

In our 
semantic information theory,
the concepts of ``proper oc-circuit,'' 
``proper logic,'' ``proper universe,'' and
``proper semantics''
introduced in Definition \ref{def:sis} 
and the computability shown in Theorem
\ref{thm:comp-sis}
represent a positive answer to the questions 
raised at the end of
Section \ref{sec:existing-semantic-theories}.

\end{remark}

We then introduce the concept of
{\it conditional oc-circuit}, 
{\it conditional semantic information 
source}
and {\it conditional semantic information 
amount},
which play central roles in our theory, 
especially in the semantic channel coding theorem
(Section \ref{sec:s-channel-coding}),
the effectiveness coding theorem 
(Section \ref{sec:effectiveness})
and in the semantic source coding theorem (Section \ref{sec:s-source-coding}).

\begin{definition}
\label{def:cond-occ} (Conditional OC-Circuits)

Let $\ZZZ \set \{Z_n\}_{n\in\N}$ be 
a sequential family of distributions.
A ``conditional oc-circuit for a family of distributions under 
$\ZZZ$'' is 
${\CCC}^{ : \ZZZ} := 
(\overline{C}^{ : \ZZZ}, \infty, 
u^{ : \ZZZ}, \vec{m}_\infty^{ : \ZZZ})$, 
where
\begin{eqnarray*}
&&
\overline{C}^{ : \ZZZ} 
\set (C^{ : \ZZZ}, N_u^{ : \ZZZ}, N_s^{ : \ZZZ}, 
N_m^{ : \ZZZ}, N_r^{ : \ZZZ}, N_z^{ : \ZZZ}, 
L_y^{ : \ZZZ}, s_1^{ : \ZZZ}),
\\
&&
\vec{m}_\infty^{ : \ZZZ} \set
(m_i)_{i=1,2,...} \set (m_1,m_2,...) \in \{0,1\}^\infty, 
\\
&&
(z_1, z_2, ...)
\rset \ZZZ 
\\
&&
\begin{tabular}{ccc}
& $\ZZZ$ & \\
& $\downarrow$ & \\ 
$(s_{i+1}, y_{i}) \longleftarrow 
\fbox{$C^{ : \ZZZ}(u^{ : \ZZZ}, \ \cdot \ )$}
\longleftarrow$ 
& $( \ \fbox{$z_i$}$ ,  & $s_{i}, m_i, r_i),$ \ 
\end{tabular}
\\
&&
\ \ \ \ \ \ \ \ \ \ \ \ \ \ \ \ \ \ 
\ \ \ \ \ \ \ \ \ \ \ \ \ \ \ \ \ \
\ \ \ \ \ \ \ \ \ \ \ \ \ \ \ 
i=1,2,...,
\\
&&
\ \ \
\mbox{i.e.,} \ 
(s_{i+1}, y_{i}) := C^{ : \ZZZ}
(u^{ : \ZZZ}, z_i, s_{i}, m_i, r_i),  \ \ i=1,2,...,
\\
&&
Y_n \set 
(y_1,..,y_{K_n})_n, \ \
K_n^{ : \ZZZ} \set \lceil n/L_y^{ : \ZZZ} \rceil
\ \ 
\\
&&
\YYY \set \{Y_n\}_{n\in\N} \rset \CCC^{ : \ZZZ},
\end{eqnarray*}
where
$m_i \in \{0,1\}^{N_m^{ : \ZZZ}}$,
$z_i \in \{0,1\}^{N_z^{ : \ZZZ}}$,  
$s_i \in \{0,1\}^{N_s^{ : \ZZZ}}$,
$u^{ : \ZZZ} \in \{0,1\}^{N_u^{ : \ZZZ}}$,
$r_i \uset \{0,1\}^{N_r^{ : \ZZZ}}$,
and
$y_i \in \{0,1\}^{L_y^{ : \ZZZ}}$.
The probability on $\YYY$ is taken over
the randomness of $\ZZZ$ and $\{r_i\}$,

\end{definition}

\begin{remark}
Given a sample of sequential family of distributions $\ZZZ$, conditional oc-circuit $\CCC^{:\ZZZ}$  divides the sample of $\ZZZ$ into $N_z^{:\ZZZ}$-bit strings, $z_1, z_2,...$, where the size, $N_z^{:\ZZZ}$,  is also determined by $\CCC^{:\ZZZ}$.

\end{remark}

%%%%%%%% 20160707

\begin{definition} 
\label{def:csia}
(Conditional Semantic Information Source and
Conditional Semantic Information Amount)

Let $\ZZZ$ be a sequential family of distributions.
A family of distributions, ${\cal X} \set \{X_n\}_{n \in \mathbb{N}}$, is called a
``conditional semantic information source
under $\ZZZ$''
at precision level $\delta(\cdot)$
if 
there exists an conditional oc-circuit for a family of distributions under $\ZZZ$,
$\CCC^{\XXX : \ZZZ} \set (\overline{C}^{\XXX:\ZZZ}, u^{\XXX:\ZZZ}, \infty, 
\vec{m}_\infty^{\XXX:\ZZZ})$,
where
$\CCC^{\XXX : \ZZZ}_n \set (\overline{C}^{\XXX:\ZZZ}, u^{\XXX:\ZZZ}, n,  
\vec{m}_n^{\XXX:\ZZZ})$
and $\vec{m}^{\XXX : \ZZZ}_n$ is the 
($N^{\XXX : \ZZZ}_m \cdot \lceil n/L_y^{\XXX : \ZZZ}\rceil$-bit)
prefix of $\vec{m}^{\XXX : \ZZZ}_\infty$
for $n$-bit output,
that satisfies the following conditions.
\begin{itemize}
\item
For all $n \in \N$, \ \ 
$
X_n \overset{\delta(n)}{\approx}
Y_n^{\XXX:\ZZZ} \rset \CCC^{\XXX:\ZZZ}_n
$
with any sampled value of $Z_{N_z^{\XXX:\ZZZ} K_n^{\XXX:\ZZZ}} \rset \ZZZ$,
and 
\item
there exists $n_0$ 
such that
for all $n > n_0$, 
\begin{eqnarray}
&&
|(\overline{C}^{\XXX:\ZZZ}, u^{\XXX:\ZZZ}, 
\vec{m}_n^{\XXX:\ZZZ})|
=
\min\{ 
|(\overline{C}^{: \ZZZ}, u^{: \ZZZ}, \vec{m}_n^{: \ZZZ})|
 \ \mid \
\nonumber
\\
&&
\ \ \ \ \ \ \ \ \ \ \ \ 
X_n \overset{\delta(n)}{\approx} 
Y_n^{ : \ZZZ} \rset 
(\overline{C}^{ : \ZZZ}, u^{ : \ZZZ}, n,  
\vec{m}_n^{ : \ZZZ})
\nonumber
\\
&&
\ \ \ \ \ \ \
\mbox{with any sampled value of $Z_{N_z^{:\ZZZ} K_n^{:\ZZZ}} \rset \ZZZ$}
 \}.
\label{eq:csia} 
\end{eqnarray}
\end{itemize}
If there are multiple conditional 
oc-circuits, $\CCC^{\XXX : \ZZZ}$, that
satisfy the above conditions, 
the lexicographically first one
is selected as $\CCC^{\XXX : \ZZZ}$.

Then, the ``conditional semantic information amount,'' 
${\SA}$, of $X_n \in \XXX$ 
under 
$\ZZZ$
at precision level $\delta(n)$ is 
$$\SA(X_n: \ZZZ,
\delta(n)) \set 
\lceil n N_m^{\XXX:\ZZZ} /L_y^{\XXX:\ZZZ}\rceil,
$$
where
$\overline{C}^{\XXX:\ZZZ} \set 
(C^{\XXX:\ZZZ}, N_u^{\XXX:\ZZZ}, N_s^{\XXX:\ZZZ},$ 
$N_m^{\XXX:\ZZZ}, N_r^{\XXX:\ZZZ}, 
N_z^{\XXX:\ZZZ},$ 
$L_y^{\XXX:\ZZZ}, s_1^{\XXX:\ZZZ}).$

\end{definition}

\begin{remark}
Eq.(\ref{eq:csia}) can be written as below in a manner similar to that for Eq.(\ref{eq:sis-2}) shown in Remark \ref{rmk:sis}.  For sufficiently large $n$, 
$$
|\CCC^{\XXX : \ZZZ}_n| 
= 
|\CCC^{X_n:\ZZZ} |,
$$
where
$\CCC^{X_n:\ZZZ} \set 
(\overline{C}^{X_n:\ZZZ}, u^{X_n:\ZZZ}, n,  
\vec{m}_n^{X_n:\ZZZ})$
is the shortest one in 
$\{ 
(\overline{C}^{: \ZZZ}, u^{: \ZZZ}, n, \vec{m}_n^{: \ZZZ})
 \ \mid \
X_n \overset{\delta(n)}{\approx} 
Y_n^{ : \ZZZ} \rset 
(\overline{C}^{ : \ZZZ}, u^{ : \ZZZ}, n, 
\vec{m}_n^{ : \ZZZ})
\ 
\mbox{with any sampled value of $Z_{N_z^{:\ZZZ} K_n^{:\ZZZ}} \rset \ZZZ$}
 \}$.

\end{remark}

Based on the conditional semantic information amount,
we next introduce the concept of the
{\it semantic mutual information amount}, 
which is employed in the 
{\it effectiveness} to be shown in
Section \ref{sec:effectiveness}.

\begin{definition} (Semantic Mutual Information Amount)

Let  
$\ZZZ \set \{Z_n\}_{n\in\N}$
be a sequential family of distributions
and
$\XXX \set \{X_n\}_{n\in\N}$ 
be a semantic information source and a conditional semantic information source
under $\ZZZ$.

The ``semantic mutual information amount''
of distribution $X_n \in \XXX$ with $\ZZZ$,
${\sf SI}(X_n: \ZZZ, \ \delta(n))$,
is defined by 
$$ {\sf SI}(X_n: \ZZZ, \ \delta(n)) \set 
\SA(X_n,\delta(n)) - \SA(X_n, \ZZZ,\delta(n)).$$ 

\end{definition}

\begin{remark}
Semantic mutual information amount
${\sf SI}(X_n: \ZZZ, \ \delta(n))$
means the semantic information amount
in $\ZZZ$ with respect to $X_n$.
More precisely, it means 
the semantic information amount 
in $Z_{\ell(n)}$ with respect to $X_n$,
where $\ell(n) \set 
N_z^{\XXX:\ZZZ}\cdot
\lceil n /L_y^{\XXX:\ZZZ}\rceil$.

In contrast to the mutual information amount in the Shannon information theory,
the semantic mutual information amount
is not symmetric, i.e.,
${\sf SI}(X_n: \ZZZ, \ \delta(n))$  
is not always equivalent to
${\sf SI}(Z_{\ell(n)}: \XXX, \ \delta(n))$  
for some $\ell(\cdot)$, since
even if some semantic information $X$ is useful for $Z$, $Z$ may not be so useful for $X$,
e.g.,
quantum physics is 
often useful to
understand chemical phenomena but 
the converse is not always true.

\end{remark}

\subsubsection{Semantic Source Coding}
\label{sec:s-source-coding}

Based on the notion of (conditional) semantic information sources,
we next develop a semantic information theory for Level B (semantic) problem as described
by Weaver \cite{Weaver49}.

Our theory answers two fundamental questions in semantic information theory: 
What is the ultimate semantic data compression, or ultimate data compression with preserving semantics, and what is the ultimate transmission rate of semantic data communication.   
The first question is answered by Theorem \ref{thm:ssc}, 
{\it semantic source coding theorem}, 
in this section, 
and the second question is answered by Theorems \ref{thm:scc1} and \ref{thm:scc2},
{\it semantic channel coding theorem}, 
in Section \ref{sec:s-channel-coding}.

To begin with, 
we introduce the notion of semantic data compression in the following definition.  

\begin{definition}\label{def:ssc}
(Semantic Source Coding System) 

Let $\XXX \set \{X_n\}_{n\in\N}$ be a semantic information source. 
Semantic source coding system 
$\SSC({\sf S},{\sf R};\XXX,\YYY,\ZZZ)$ consists of
sender ${\sf S}$,
which outputs a sequential family of distributions, $\ZZZ$,  
on 
$\XXX$,   
and receiver ${\sf R}$, which is  
a conditional oc-circuit $\CCC^{ : \ZZZ}$ 
under $\ZZZ$
without semantics input ($N_m^{ : \ZZZ}=0$)
to output a family of distributions 
$\YYY$, where
${\CCC}^{ : \ZZZ} := 
(\overline{C}^{ : \ZZZ}, u^{ : \ZZZ}, \infty, \vec{m}_\infty^{ : \ZZZ})$,
and
$\overline{C}^{ : \ZZZ} 
\set (C^{ : \ZZZ}, N_u^{ : \ZZZ}, N_s^{ : \ZZZ}, 
N_m^{ : \ZZZ}, N_r^{ : \ZZZ}, N_z^{ : \ZZZ}, 
L_y^{ : \ZZZ}, s_1^{ : \ZZZ})$.

For parameter $n \in \N$,
sender ${\sf S}$ outputs   
$Z_{\ell(n)}$ on $X_n \in \XXX$,
which is directly input to receiver ${\sf R}$,
and
${\sf R}$ outputs 
$Y_n \rset {\sf R}(n,Z_{\ell(n)})$,
where 
$\ell(n) \set N_z^{ :\ZZZ} \cdot  
\lceil n/L_y^{ : \ZZZ} \rceil$, 
$Z_{\ell(n)} \in \ZZZ$ is a distribution  
over $\{0,1\}^{\ell(n)}$, and
$Y_n \in \YYY$ is a distribution over 
$\{0,1\}^n$.

\begin{center}
\begin{tabular}{cccccc}
 & $Z_{\ell(n)}$ 
 & \\
 \fbox{${\sf S}^{\XXX}(n)$} & $\longrightarrow$ & 
 \fbox{${\sf R}(n,\cdot)$}  $\ra Y_n$ 
\end{tabular}
\end{center}

We call $\ell(n)$ the code length of 
$\SSC({\sf S},{\sf R};\XXX,\YYY,\ZZZ)$ in $n \in \N$. 

We say that $\SSC({\sf S},{\sf R};\XXX,\YYY,\ZZZ)$ correctly 
codes at precision level $\delta(\cdot)$ 
if there exists $n_0\in \mathbb{N}$ 
such that
for all $n \geq n_0$, \ 
$X_n \overset{\delta(n)}{\approx} Y_n$
with any sampled value of $Z_{\ell(n)} \rset \ZZZ$. 

\end{definition}

The following theorem answers the above-mentioned first question. 
Roughly, the ultimate compression size of semantic information source $\XXX \set \{X_n\}_{n\in\N}$ at precision level $\delta(\cdot)$ is its semantic information amount, $\SA(X_n, \delta(n))$, asymptotically.   
The compression size of $\SA(X_n, \delta(n)) + \epsilon$ ($\epsilon$ is a positive small value) is possible, but the compression shorter than $\SA(X_n, \delta(n)) - \epsilon$ is impossible.
   
\begin{theorem}\label{thm:ssc} (Semantic Source Coding Theorem)

Let ${\cal X}\set \{X_n\}_{n\in\N}$ be a semantic information source with precision level 
$\delta(\cdot)$. 

There exists a semantic source coding system 
$\SSC({\sf S},{\sf R};{\cal X},\YYY,\ZZZ)$
with code length $\ell(\cdot)$
that correctly codes at precision level $\delta(\cdot)$ 
and 
$\ell(\cdot)$ satisfies the following
inequality. 
For any $\epsilon > 0$ 
there exists $n_0 \in \N$
such that 
for all $n > n_0$, 
$$\SA(X_n, \delta(n)) \leq \ell(n)
< \SA(X_n, \delta(n)) + \epsilon. 
$$

There is no semantic source coding system $\SSC({\sf S},{\sf R};{\cal X},\YYY,\ZZZ)$
such that
$\SSC({\sf S},{\sf R};{\cal X},\YYY,\ZZZ)$
with code length $\ell(\cdot)$
correctly codes at precision level $\delta(\cdot)$ and
$\ell(\cdot)$ satisfies the following
inequality. 
For any $\epsilon >0$, there exists $n_0\in\N$ for all $n > n_0$,
$$ \ell(n) < \SA(X_n, \delta(n)) -\epsilon.$$

\end{theorem}

\noindent
{\bf Proof}

Since ${\cal X}\set \{X_n\}_{n\in\N}$ is 
a semantic information source,
the proper oc-circuit of ${\cal X}$, 
$\CCC^{\XXX} \set (\overline{C}^{\XXX}, u^{\XXX}, \infty,  \vec{m}^{\XXX}_\infty)$,
exists,
where ${\CCC}^{\XXX}_n := 
(\overline{C}^{\XXX}, u^{\XXX}, n,  \vec{m}^{\XXX}_n)$
and
$\overline{C}^{\XXX}
\set 
(C^{\XXX}, N_u^{\XXX}, N_s^{\XXX}, N_m^{\XXX}, N_r^{\XXX}, 
L_y^{\XXX}, s_1^{\XXX})$.

Then,
$\SA(X_n,\delta(n)) \set 
\lceil nN_m^{\XXX}/L_y^{\XXX}\rceil$ \ 
($\approx 
\lceil n/L_y^{\XXX}\rceil N_m^{\XXX} 
= |\vec{m}_n^{\XXX}|)$,
where 
$\SA(X_n,\delta(n))\approx 
|\vec{m}_n^{\XXX}|$ means
$\forall \epsilon > 0 \ \exists n_0 \ \forall n > n_0 \
\SA(X_n,\delta(n))) \leq 
|\vec{m}_n^{\XXX}| < \SA(X_n,\delta(n))) + \epsilon$.

We construct semantic source coding system 
$\SSC({\sf S},{\sf R};\XXX,\YYY,\ZZZ)$ 
such that
sender ${\sf S}$ sends $\vec{m}_n^{\XXX}$ 
(as ${Z}_{\ell(n)}$) to
receiver ${\sf R}$, i.e.,
$\ell(n) \set |\vec{m}_n^{\XXX}|
\approx \SA(X_n,\delta(n))$,
and ${\sf R}$ is a conditional oc-circuit under $\ZZZ$ without semantics input,
${\CCC}^{ : \ZZZ} \set 
(\overline{C}^{ : \ZZZ}, u^{ : \ZZZ}, \infty, \lambda)$, 
where
$u^{ : \ZZZ} \set u^{\XXX}$,
and
$\overline{C}^{ : \ZZZ} 
\set (C^{\XXX}, 
N_u^{\XXX}, N_s^{\XXX}, 
0, 
N_r^{\XXX}, N_m^{\XXX},
L_y^{\XXX}, 
 s_1^{\XXX})$
as
$(C^{ : \ZZZ}, N_u^{ : \ZZZ}, N_s^{ : \ZZZ}, 
N_m^{ : \ZZZ}, N_r^{ : \ZZZ}, N_z^{ : \ZZZ}, 
L_y^{ : \ZZZ}, s_1^{ : \ZZZ})$.
Here,
${\overline C}^{ : \ZZZ}$ is the same functionality as 
that of ${\overline C}^{\XXX}$ except the input place such that
$\vec{m}_n^{\XXX}$ sent from
${\sf S}$ is input to ${\overline C}^{ : \ZZZ}$
as ${Z}_{\ell(n)}$, i.e., 
$N_z^{ : \ZZZ} \set N_m^{\XXX}$,
and no semantics is input to ${\overline C}^{ : \ZZZ}$, 
i.e., $N_m^{ : \ZZZ} \set 0$, 
while $\vec{m}_n^{\XXX}$ is input to
${\overline C}^{\XXX}$ as the semantics.

From the definition of the proper oc-circuit,
for sufficiently large $n$ ($\exists n_0 \ \forall n > n_0$),
$$
{X}_n \overset{\delta(n)}{\approx} 
Y_n \rset 
({\overline C}^{\XXX}, u^{\XXX}, n,\vec{m}_n^{\XXX})
=
({\overline C}^{ : \ZZZ}, u^{\XXX}, n,  \lambda).
$$
That is, for sufficiently large $n$,
$
{X}_n \overset{\delta(n)}{\approx} 
Y_n \rset  
{\sf R}(n, Z_{\ell(n)})
$, i.e.,
the constructed $\SSC({\sf S},{\sf R};\XXX,\YYY,\ZZZ)$ correctly 
codes at precision level $\delta(\cdot)$.

Since
$\ell(n) 
\approx \SA(X_n,\delta(n))$, \ \ \
$\forall \epsilon > 0 \ \exists n_0 \ \forall n > n_0$
$$
\SA(X_n,\delta(n))) \leq 
\ell(n)  < \SA(X_n,\delta(n))) + \epsilon.
$$
This completes the former statement of this theorem.

To prove the latter statement of this theorem by contradiction,  
let us assume that there exists semantic source coding system
$\SSC({\sf S},{\sf R};{\cal X},\YYY,\ZZZ)$
such that
$\SSC({\sf S},{\sf R};{\cal X},\YYY,\ZZZ)$
correctly codes at precision level $\delta(\cdot)$ and
its code length $\ell(n)$ is that 
$\forall \epsilon >0$, $\exists n_0\in\N$, $\forall n > n_0$,
%for any $\epsilon >0$, there exists $n_0\in\N$ for all $n > n_0$,
$ \ell(n) < \SA(X_n, \delta) -\epsilon.$

Since $\ZZZ$ is a sequential family of distributions and 
$\SSC({\sf S},{\sf R};{\cal X},\YYY,\ZZZ)$
correctly codes at precision level $\delta(\cdot)$ 
with any sampled value of $\ZZZ$,
there exists $z_{\ell(n)} \in \{0,1\}^{\ell(n)}$ such that
$z_{\ell(n)}$ is the $\ell(n)$-bit prefix of
a sampled value of $\ZZZ$,
and 
for sufficiently large $n$,
$
{X}_n \overset{\delta(n)}{\approx} 
Y_n \rset  
{\sf R}(n, z_{\ell(n)})
$.

As shown in the proof of the former statement, 
${\sf R}(n, z_{\ell(n)})$, i.e., conditional oc-circuit ${\overline C}^{ : \ZZZ}$ with a sampled value
of $\{z_{\ell(n)}\}$ and no semantics input 
is the same functionality as 
oc-circuit ${\overline C}$ 
with semantics input $\{z_{\ell(n)}\}$.
That is, there exists an oc-circuit
$({\overline C}, u, n, z_{\ell(n)})$ such that
for sufficiently large $n$,
$$
{X}_n \overset{\delta(n)}{\approx} 
Y_n \rset  
({\overline C}, u, n, z_{\ell(n)}).
$$
Since for any $\epsilon$ and sufficiently large $n$ \  
$ \epsilon < \SA(X_n, \delta) - \ell(n)$ and 
$|({\overline C}, u)|$ is bounded,
it holds that for sufficiently large $n$ 
$$
|({\overline C}, u) | - |({\overline C}^{\XXX}, u^{\XXX})| 
< \SA(X_n, \delta) - \ell(n)
\leq |\vec{m}^{\XXX}_n| - |z_{\ell(n)}|,
$$
where $({\overline C}^{\XXX}, u^{\XXX}, \vec{m}^{\XXX}_n)$ is the proper oc-circuit of $\XXX \set \{X_n\}_{n\in\N}$.
That is,
$$
|({\overline C}, u, z_{\ell(n)})|
< |({\overline C}^{\XXX}, u^{\XXX}, \vec{m}^{\XXX}_n)|. 
$$

It contradicts the minimality of $|({\overline C}^{\XXX}, u^{\XXX}, \vec{m}^{\XXX}_n)|$ and completes the proof of the latter statement.

\qed
\vspace{5pt}

\begin{remark}
If semantic information source $\XXX \set \{X_n\}_{n\in\N}$ is a family of uniform distributions, its semantic information amount is zero, or ${\sf SA}(X_n, \delta(n)) = 0$ for any $n \in \N$ and $\delta(\cdot)$.  
Hence, semantically compressed data size, $\ell(n)$, of $X_n$ can be almost 0, due to Theorem \ref{thm:ssc}.

It is highly contrast to the data compression capability in the traditional (Shannon) information theory:  the above-mentioned source, $\XXX$, cannot be compressed any more because its Shannon entropy is the maximum. 

As shown in this example, the semantic data compression should be more capable than the traditional data compression in many applications.
That is, the semantic data compression indicated by Theorem \ref{thm:ssc} offers a great potential in various practical applications of data compression.

\end{remark}

\subsubsection{Semantic Channel Coding} 
\label{sec:s-channel-coding} 
 
In this section, we answer the second question described in the beginning of Section \ref{sec:s-source-coding}:
What is the ultimate transmission rate of semantic data communication.  

First, in the following definition, we introduce a model of 
%a coding in the semantic information space and
semantic communication channel. %and semantic channel coding system. 
Here, the semantic communication channel may be noisy or the received data from the channel may contain semantic errors. Various types of semantics errors or noises are investigated in \cite{BaoBasDeaParSwaLel11}.   
The notion of a semantic channel coding system is introduced to correct such errors in the semantic information space over the communication channel.

\begin{definition} 
\label{def:scc}
(Semantic Channel Coding System)

Semantic channel coding system 
$\SCC \set 
\SCC({\sf S}^{(\overline{C},u,{\sf code})},{\sf R}, {\sf Ch};\XXX,\YYY,\ZZZ)$ 
consists of sender ${\sf S}$,
which has a coding machine, 
$(\overline{C},u,{\sf code})$,
and outputs 
a sequential family of distributions, $\XXX$;
communication channel $\sf Ch$, which
receives $\XXX$ and outputs 
a sequential family of distributions,
$\ZZZ$;
and 
receiver ${\sf R}$, 
which is a conditional oc-circuit 
under $\ZZZ$
and outputs $\YYY$.

Here, $\XXX$ is generated by 
an oc-circuit, ${\CCC} := 
(\overline{C}, u, \infty,  
\vec{m}_\infty)$, 
where
$\overline{C} \set 
({C}, N_u, N_s, N_m, N_r,L_y,s_1)$ 
(logic),   
$u \in \{0,1\}^{N_u}$
(universe space), 
$\vec{m}_\infty \in \mathbb{M}^{\overline{C}} \set 
\{ (m_i)_{i=1,2,..} \ \mid \  
m_i \in \{0,1\}^{N_m} \}$
(semantic information space),
and  
$\mathbb{M}^{\overline{C}}_n 
= \{0,1\}^{N_m K_n}$ 
(the $N_m K_n$-bit prefix of $\mathbb{M}^{\overline{C}}$ for $n$-bit output) 
($K_n \set \lceil n/L_y \rceil$).
A coding, ${\sf code}$, with $n$ is 
${\sf code}_n: \{0,1\}^{k(n)} \ra \mathbb{M}^{\overline{C}}_n$.

For parameter $n \in \N$,
given  
$\vec{m}^+_{n} \in 
\{0,1\}^{k(n)}$, 
${\sf S}$ computes
$\vec{m}_{n} \set {\sf code}_n(\vec{m}^{+}_{n}) \in \mathbb{M}^{\overline{C}}_n$
and
$X_n 
\rset 
(\overline{C}, u, n, \vec{m}_{n})$,
where
$X_n \in \XXX$ is a distribution 
over $\{0,1\}^n$.
$X_n$ is input to channel $\sf Ch$,
and 
$\sf Ch$ 
outputs $Z_{\ell(n)} \in \ZZZ$,
where $Z_{\ell(n)}$ is a distribution 
over $\{0,1\}^{\ell(n)}$.
Receiver ${\sf R}$ (conditional oc-circuit under $\ZZZ$) receives $Z_{\ell(n)}$ 
and outputs $Y_n \in \YYY$,
where $Y_n$ is a distribution over
$\{0,1\}^n$.

\begin{center}
\begin{tabular}{cccccc}
 & $X_n$ & & $Z_{\ell(n)}$ & \\
 \fbox{${\sf S}^{(\overline{C},u, 
{\sf code})}(n,\vec{m}^+_n)$} & $\longrightarrow$ & 
 \fbox{${\sf Ch}(n,\cdot)$} & $\longrightarrow$ & 
 \fbox{${\sf R} 
 (n,\cdot)$}  $\ra Y_n$ \\
\end{tabular}
\end{center}

We say that 
$\SCC \set \SCC({\sf S}^{(\overline{C},u,{\sf code})},{\sf R}, {\sf Ch};\XXX,\YYY,\ZZZ)$ correctly 
codes at precision level $\delta(\cdot)$ 
if there exists $n_0\in \mathbb{N}$ 
such that
for all $n \geq n_0$, \ 
for all $\vec{m}^+_{n} \in 
\{0,1\}^{k(n)}$, \ 
$X_n \overset{\delta(n)}{\approx} Y_n$
with any sampled value of $Z_{\ell(n)} \rset \ZZZ$.

\end{definition}

We next define the {\it semantic channel capacity} of a semantic channel and {\it semantic communication rate} of a semantic coding.
Roughly, the semantic channel capacity represents the maximum rate of semantic information that can be transmitted over the semantic channel, or the ratio of the maximum semantic communication amount (size) to the communication data size. 
The semantic communication rate is the ratio of semantic information size (input size to the coding) to the communication data size, $k(n)/n$, in the semantic coding. 

\begin{definition}\label{def:scc-scr} 
(Semantic Channel Capacity and 
Semantic Communication Rate)

Let 
$\SC \set ({\sf S}^{(\overline{C},u)}, {\sf Ch};\XXX,\ZZZ)$
be a ``semantic channel'' of semantic channel coding system 
$\SCC({\sf S}^{(\overline{C}, u,{\sf code})},$ ${\sf R}, {\sf Ch};\XXX,\YYY,\ZZZ)$.
%where $\overline{C} \set (C, N_u, N_s, N_m, N_r, L_x, s_1)$, $u$, and $\mathbb{M}^{\overline{C}}$ are the logic, universe, and semantic information space of an oc-circuit for a family of distributions, respectively. 
Given $\vec{m}_\infty \in \mathbb{M}^{\overline{C}}$,
${\sf S}$ in $\SC$ computes
$\XXX \set \{X(\vec{m}_n)
\set X_n \rset (\overline{C}, u,n,\vec{m}_n) \}_{n \in \N}$, i.e., $\XXX \rset (\overline{C}, u, \infty, \vec{m}_\infty)$, where $\vec{m}_n$ is the 
$N_m \cdot \lceil n/L_y \rceil$-bit
prefix of 
$\vec{m}_\infty$ for $n$-bit output. 
Then, $\XXX \set \{X(\vec{m}_n)\}_{n \in \N}$ is input to channel $\sf Ch$ in $\SC$,
and 
$\sf Ch$ 
outputs $\ZZZ \set \{Z_{\ell(n)}\}_{n \in \N}$.

If 
for any $\vec{m}_\infty \in \mathbb{M}^{\overline{C}}$, 
$\XXX \rset (\overline{C}, u, \infty, \vec{m}_\infty)$ is a conditional semantic information source under $\ZZZ$ in semantic channel $\SC$, 
we call $\SC$ ``{\it normal}''.

Let $\MMM \set \{M_n\}_{n \in \N}$ be a sequential family of distributions 
over $\mathbb{M}^{\overline{C}}$,
where
$M_n$ is a distribution 
over 
$\mathbb{M}^{\overline{C}}_n$,  
$n\in\N$, i.e., 
$M_n := \{ (\vec{m}_n, p_{\vec{m}_n}) 
\mid \vec{m}_n \in \mathbb{M}^{\overline{C}}_n\}$,
(see Definition \ref{def:seq-fd}
for the sequential family of distributions).

When semantic channel $\SC$ is normal,  
``semantic channel capacity'' ${\sf SC}$ of  
$\SC$ 
for $n \in \N$
(say $\SC_n$)
is
\begin{eqnarray}
&&
{\sf SC}(\SC_n,
\delta(n)) 
\set 
\nonumber
\\
&& \
\frac{1}{n}\cdot 
\max_{M_n \in \MMM^{\sf seq}_n}\{
H(M_n) 
- \ {\sf E}_{M_n} 
(\SA(X(\vec{m}_n): \ZZZ, \delta(n))
\},\ \ \ \ \
\label{eq:sc0}
\end{eqnarray}
where 
$\MMM^{\sf seq}_n$ 
is the class of 
the $N_m \cdot \lceil n/L_y \rceil$-bit
prefix of sequential families of distributions,
i.e., $M_n \in \MMM^{\sf seq}_n$
is the $N_m \cdot \lceil n/L_y \rceil$-bit
prefix of a sequential family of distributions,
$H(\cdot)$ is the Shannon entropy and 
${\sf E}_{M_n}  
(\cdot)$
is the expectation value over the distribution of
$\vec{m}_n \rset M_n \in \MMM^{\sf seq}_n$.

``Semantic communication rate'' ${\sf SR}$ of 
semantic coding $(\overline{C},u, {\sf code})$ 
in semantic channel coding system 
$\SCC({\sf S}^{(\overline{C}, u,{\sf code})},$ ${\sf R}, {\sf Ch};\XXX,\YYY,\ZZZ)$ for $n \in \N$ is
\begin{eqnarray}
{\sf SR}({\sf code}_n) \set 
k(n)/n. \ \ \  
\end{eqnarray}

We say a semantic channel coding system, $\SCC$, is ``normal,'' if the semantic channel, $\SC$, of $\SCC$ is normal.

\end{definition}

We now show that the semantic capacity is the upper limit (or theoretically maximum) rate of semantic data transmission at which we can send semantic information over the semantic channel and recover the information at the output in the semantic channel coding system.

\begin{theorem} \label{thm:scc1}(Semantic Channel Coding Theorem (1))

There exists no normal semantic channel coding system 
$\SCC({\sf S}^{(\overline{C}, u,{\sf code})},$ ${\sf R}, {\sf Ch};\XXX,\YYY,\ZZZ)$
that correctly 
codes at precision level $\delta(\cdot)$
and 
there exists $n_0\in\N$ such that 
for all $n > n_0$, \  
$$ 
{\sf SC}(\SC_n,
\delta(n)) < {\sf SR}({\sf code}_n). 
$$

\end{theorem}

\noindent
{\bf Proof}

To prove this theorem by contradiction,
we first assume that 
$\SCC$ correctly codes at precision level $\delta(\cdot)$
while satisfying  
$
{\sf SR}({\sf code}_n) > 
{\sf SC}(\SC_n,\delta(n)). 
$

We then construct a distribution, 
$M^+_n := \{ (\vec{m}_n, p_{\vec{m}_n})\}$, 
such that
$p_{\vec{m}_n}\set 1/2^{k(n)}$
if $\vec{m}_n \set {\sf code}_n(\vec{m}^+_n)$ with $\vec{m}^+_n \in \{0,1\}^{k(n)}$, \
and 
$p_{\vec{m}_n}\set 0$
\ otherwise.
Clearly,  
$H(M^+_n) = k(n) = n\cdot {\sf SR}({\sf code}_n)$.

Since $\SCC$ correctly 
codes at precision level $\delta(\cdot)$
for any $\vec{m}_n \set {\sf code}_n(\vec{m}^+_n)$ which occurs with probability 1
in $M^+_n$,
$\sf R$ of $\SCC$, a conditional oc-circuit under $\ZZZ$, outputs $Y_n \overset{\delta(n)}{\approx}
X(\vec{m}_n)$.  
That is,
$\SA(X(\vec{m}_n): \ZZZ, \delta(n)) = 0$
for any $\vec{m}_n$ that occurs in 
$M^+_n$ with non-zero probability.
Therefore,
${\sf E}_{M^+_n} 
(\SA(X(\vec{m}_n): \ZZZ, \delta(n))
=0$.

Due to the definition (maximality) of ${\sf SC}(\SC_n,\delta(n))$, we obtain  that
${\sf SC}(\SC_n,\delta(n)) 
\geq 
\frac{1}{n}\cdot 
(H(M^+_n)
- \ {\sf E}_{M^+_n} 
(\SA(X(\vec{m}_n): \ZZZ, \delta(n))
=$
$\frac{1}{n}\cdot
(n\cdot {\sf SR}({\sf code}_n))$
$= {\sf SR}({\sf code}_n)$.

This contradicts the assumption that
$
{\sf SR}({\sf code}_n) > 
{\sf SC}(\SC_n,\delta(n)). 
$.

\qed
\vspace{5pt}

In the theorem above, it is shown that a semantic data transmission rate over a channel is impossible beyond the semantic capacity of the channel.
We next present that a semantic transmission rate slightly below the semantic capacity is possible over a class of semantic channels,
{\it uniform} semantic channels.      
%First, we introduce the class of uniform semantic channels.      

\begin{definition}
\label{def:usc}
(Uniform Semantic Channel)

Let
$M^*_n$ be the value (distribution) of $M_n \in \MMM^{\sf seq}_n$ to maximize  
Eq.(\ref{eq:sc0}) and 
$\{ (\vec{m}_n, p^*_{\vec{m}_n}) 
\mid \vec{m}_n \in \mathbb{M}^{\overline{C}}_n\}
\set M^*_n$.

We say a normal semantic channel, $\SC$,
is ``uniform'' 
if the following conditions hold.
\begin{itemize}
\item 
(Consistency)
$\MMM^* \set \{M^*_n\}_{n\in \N}$
is a sequential family of distributions.

\item
(Uniformity of $\MMM^*$) 

For any $\epsilon_1, \epsilon_2>0$, there exists $n_0 \in \N$
such that for all $n > n_0$
\begin{eqnarray*}
&&
\Pr
[ \ | -\log_2{p^*_{\vec{m}_n}} - H(M^*_n)| > \epsilon_1
] < \epsilon_2,
\end{eqnarray*}
where the probability is taken over the 
randomness of $\vec{m}_n \rset M^*_n$.

\item
(Uniformity of conditional ${\sf SA}$)

For any $\epsilon_1, \epsilon_2>0$, there exists $n_0 \in \N$
such that for all $n > n_0$
\begin{eqnarray*}
&&
\Pr
[ |\SA(X(\vec{m}_n):\ZZZ,\delta(n)) - T^*| > \epsilon_1 
] < \epsilon_2,
\end{eqnarray*}
where
the probability is taken over the 
randomness of $\vec{m}_n \rset M^*_n$, and
$T^* \set
{\sf E}_{M^*_n} 
(\SA(X(\vec{m}_n):\ZZZ,\delta(n))$.

\end{itemize}

We say a semantic channel coding system, $\SCC$, is ``uniform,'' if the semantic channel, $\SC$, of $\SCC$ is uniform.

\end{definition}

\begin{theorem} \label{thm:scc2}(Semantic Channel Coding Theorem (2))

There exists a uniform semantic channel coding system, 
$\SCC({\sf S}^{(\overline{C}, u,{\sf code})},$ ${\sf R}, {\sf Ch};\XXX,\YYY,\ZZZ)$,
that correctly codes 
at precision level $\delta(\cdot)$,
and 
for any $\epsilon \ (0 <\epsilon)$, 
there exits $n_0\in\N$ such that 
for all $n > n_0$, \ 
\begin{eqnarray*}
&&
{\sf SC}(\SC_n,\delta(n)) - \epsilon 
< 
{\sf SR}({\sf code}_n) 
<
{\sf SC}(\SC_n,\delta(n)). 
\end{eqnarray*}

\end{theorem}

\noindent
{\bf Proof}

Let $M^*_n$ be the maximum value of $M_n$ 
to maximize Eq.(\ref{eq:sc0}) of  
normal channel ${\sf Ch}$ on 
$(\overline{C},u,\mathbb{M}^{\overline{C}})$
as described in Definition \ref{def:scc-scr}. 

For $\epsilon_1 >0$ and $M_n^*$, let  
$\MMM^*_{\epsilon_1, n} \set \{\mu  \mid \
(\mu, p^*_{\mu}) \in M_n^* \ \land \ | -\log_2{p^*_{\mu}} - H(M^*_n)| \leq \epsilon_1 \
\land |\SA(X(\mu):\ZZZ,\delta(n)) - T^*| \leq \epsilon_1 
\}.$

From the uniformity of $\MMM^*$ and
uniformity of the conditional SA,
it holds that 
$\forall \epsilon_1, \epsilon_2 > 0$, \
$\exists n_0\in\N$, \ 
$\forall n > n_0$, \
$\Pr[\mu \in \MMM^*_{\epsilon_1, n} \mid
 \mu \rset M_n^*] > 1 - \epsilon_2$. 
Therefore,
$\forall \epsilon_1, \epsilon_2 > 0$, \
$\exists n_0\in\N$, \ 
${\#}\MMM^*_{\epsilon_1, n} 
> 2^{H(M^*_n) - \epsilon_2}$.

%Since $|(\overline{C}^{\XXX:\ZZZ}, u^{\XXX:\ZZZ}, \vec{m}_n^{\XXX:\ZZZ})|$ equals the minimum value of $\{ |(\overline{C}^{: \ZZZ}, u^{: \ZZZ}, \vec{m}_n^{: \ZZZ})| \ \mid \ X_n \overset{\delta(n)}{\approx} Y_n^{ : \ZZZ} \rset (\overline{C}^{ : \ZZZ},$ $u^{ : \ZZZ},$ $n,$  $\vec{m}_n^{ : \ZZZ})$ with any sampled value of$Z_{N_z^{:\ZZZ} K_n^{:\ZZZ}} \rset \ZZZ$ $\}$ for sufficiently large $n$ (Eq.(\ref{eq:csia}) in Definition \ref{def:csia}), for each $\mu \in {\MMM}^*_{\epsilon_1, n}$,

Due to the definition of conditional semantic information amount $\SA(X(\mu):\ZZZ,\delta(n))$,
there exist at most $2^{\SA(X(\mu):\ZZZ,\delta(n))}$ 
distinct values of $\mu^- \in \mathbb{M}^{\overline{C}^{\XXX:\ZZZ}}_n$ such that
the output of ${\sf Ch}(n,X(\mu))$
is indistinguishable from that of ${\sf Ch}(n,X(\mu^-))$.
Let ${\cal I}_\mu \subseteq \mathbb{M}^{\overline{C}^{\XXX:\ZZZ}}_n$ be the set of such at most $2^{\SA(X(\mu):\ZZZ,\delta(n))}$ 
values of $\mu^-$.

If ${\cal I}_{\mu} \cap {\cal I}_{\mu'} \not= \lambda$
for $\mu \not= \mu'$ (the intersection of the two sets is not empty), 
there exists $\mu^- \in {\cal I}_{\mu} \cap {\cal I}_{\mu'}$ such that 
the output of ${\sf Ch}(n,X(\mu^-))$
is indistinguishable from both ${\sf Ch}(n,X(\mu))$
and ${\sf Ch}(n,X(\mu'))$, i.e.,
the output of ${\sf Ch}(n,X(\mu))$
is indistinguishable from that of ${\sf Ch}(n,X(\mu'))$.
That is,
${\cal I}_{\mu} = {\cal I}_{\mu'}$.
Therefore, for $\mu \not= \mu'$, 
either ${\cal I}_{\mu} \cap {\cal I}_{\mu'} = \lambda$
or ${\cal I}_{\mu} = {\cal I}_{\mu'}$.
Hence, %$\mathbb{M}^{\overline{C}^{\XXX:\ZZZ}}_n$ is divided into 
we have $t$ disjoint sets ${\cal I}_{\mu_1}$,
${\cal I}_{\mu_2}$, ..., ${\cal I}_{\mu_t}$ for some $t \in \N$. 
Therefore, from the above property,
${\MMM}^*_{\epsilon_1,n}$ is divided into 
disjoint equivalence classes,
${\cal I}^*_i 
\set {\cal I}_{\mu_i} \cap {\MMM}^*_{\epsilon_1,n}$,
$i=1, .., t$.

Since  $|\SA(X(\mu_i):\ZZZ,\delta(n)) - T^*| \leq \epsilon_1$ for $\mu_i \in {\MMM}^*_{\epsilon_1,n}$, 
we obtain
${\#}{\cal I}_{\mu_i} \leq 2^{\SA(X(\mu_i):\ZZZ,\delta(n))} \leq 2^{T^*+\epsilon_1}$.
Hence, 
$t %\set {\#}\{{\cal I}^*_i \mid i=1,..,t\}
\geq$ ${\#}{\MMM}^*_{\epsilon_1,n} / 
\max\{{\#}{\cal I}_{\mu_i}\}$
$\geq 2^{H(M^*_n) - \epsilon_1} /
2^{T^*+\epsilon_1}$
$= 2^{H(M^*_n) - T^* - 2\epsilon_1}$.
Let $t^* \set 2^{H(M^*_n) - T^* - 2\epsilon_1}$.

We now set a coding with $n$,
${\sf code}_n: \{0,1\}^{k(n)} \ra 
\MMM_n^{\overline{C}}$,
such that
${\sf code}_n: i \mapsto \mu_i\in {\cal I}^*_i$
for $i =1, .., t^*$.
That is,
$k(n) \set \log t^* = H(M^*_n) - T^* - 2\epsilon_1$, i.e.,
${\sf SR}({\sf code}_n) \set k(n)/n = (H(M^*_n)-T^* - 2\epsilon_1)/n$
$= {\sf SC}(\SCC_n,\delta(n)) -  2\epsilon_1/n$,
i.e.,
${\sf SR}({\sf code}_n) = {\sf SC}(\SCC_n,\delta(n)) 
- 2\epsilon_1/n$.

Since 
${\cal I}^*_i (i=1,..,t^*)$
are disjoint and 
${\sf Ch}(n,X(\mu_i)) (i=1,..,t^*)$ 
are distinct, 
the coding by ${\sf code}_n$ can be uniquely decoded.
 
Thus,
for any $\epsilon$ %(\set 2\epsilon_1)$ 
there exists $n_0$ such that
for all $n > n_0$ \  
the coding system satisfies   
${\sf SC}(\SCC_n,\delta(n)) - \epsilon 
< 
{\sf SR}({\sf code}_n) 
<
{\sf SC}(\SCC_n,\delta(n)).$ 
This completes the proof of this theorem.

\qed
\vspace{5pt}

\begin{remark}
The error correction techniques based on the traditional (Shannon) information theory have been widely used in many applications, but they are incompetent for correcting various types of semantic errors which are described in \cite{BaoBasDeaParSwaLel11}.

In this paper, the notion of a semantic channel coding system or semantic error correction is introduced. %in the semantic information space. 
Theorem \ref{thm:scc2} shows a great potential of 
semantic error correction techniques. %to solve these issues.
For example, we can correct semantic errors in a natural language sentence using the semantics and context. Such a capability of human beings is theoretically captured and formalized in Theorem \ref{thm:scc2}. Unfortunately, the proof of the theorem ignores the efficiency and the error-correction technique used in the proof is impractical, i.e., it  only gives a theoretical feasibility. However, such a feasibility result indicated by Theorem \ref{thm:scc2} should  push toward developing practical techniques. This is similar to that where Shannon's channel coding theorem is only a feasibility result and many practical error correcting codes have been developed which are quite different from Shannon's random coding technique. Towards practical semantic error correcting code techniques, artificial intelligence technologies might offer some potent means to yield a breakthrough in this field.  

\end{remark}

\subsubsection{Effectiveness Problem}
\label{sec:effectiveness}

We now consider the problem of effectiveness (Level C problem given by Weaver \cite{Weaver49} as introduced in Section \ref{sec:bacgroud}) 
in our semantic information theory.

First consider the following experiment. 
Person $S$ requests robot $R$, 
e.g., by voice using English sentences,
to perform a series of actions, and
device $D$ detects the image of the actions 
of robot $R$ and outputs the (digital form of) video image.

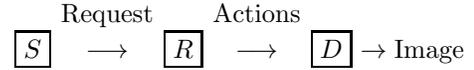
\begin{center}
\begin{tabular}{cccccc}
 & {\rm Request} & & {\rm Actions} & \\
 \fbox{$S$} & $\longrightarrow$ & 
 \fbox{$R$} & $\longrightarrow$ & 
 \fbox{$D$} $\ra 
 {\rm Image} 
 $ 
\end{tabular}
\end{center}

\

In the experiment, then, an evaluator, e.g., a human, 
$E$ compares the request and the image, 
and   
evaluates how correctly the robot understood the request and performed the
requested actions.

The problems here 
are that the request and image are
different types of information,
e.g., the request is the (digital form) voice speaking English sentences
and the image is the (digital form) video
images of the robot's actions.
Although they are different forms of 
information,
human $E$ can evaluate the capability of robot $R$
since $E$ knows some semantics common between 
English requests and the robot's actions.  

Our semantic information theory can treat such 
semantics that are common between different forms.

\begin{definition}\label{def:eff} (Effectiveness)

Message-to-conduct system
${\sf M2C}({\sf S},{\sf R},{\sf D};\XXX,\ZZZ)$ 
consists of sender ${\sf S}$
and receiver ${\sf R}$
where
${\sf S}$ 
sends semantic information source $\XXX$
to ${\sf R}$ and ${\sf R}$'s conduct is observed by some device, ${\sf D}$, which outputs a sequential family of distributions, $\ZZZ$.
 
For parameter $n \in \N$,
message $X_n \in \XXX$, which is 
the distribution over $\{0,1\}^n$,
is given to ${\sf R}$
and 
${\sf R}$'s conduct is observed as 
distribution $Z_{\ell(n)}$,
which is the distribution 
over $\{0,1\}^{\ell(n)}$.

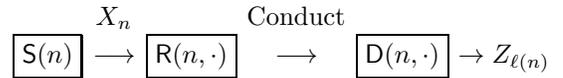
\begin{center}
\begin{tabular}{cccccc}
 & $X_n$ & & {\rm Conduct} & \\
 \fbox{${\sf S}(n)$} & $\longrightarrow$ & 
 \fbox{${\sf R}(n,\cdot)$} & $\longrightarrow$ & 
 \fbox{${\sf D}(n,\cdot)$} $\ra Z_{\ell(n)}$ \\
\end{tabular}
\end{center}

If $\XXX$ is a conditional semantic information source under $\ZZZ$,
we can define the concept of 
``effectiveness'' as follows:

The effectiveness, ${\sf Eff}(X_n: \ZZZ, \delta(n))$, 
of message-to-conduct system
${\sf M2C}({\sf S},{\sf R},{\sf D};\XXX,\ZZZ)$
is   
$$
{\sf Eff}(X_n: \ZZZ, \delta(n)) \set
\frac{{\sf SI}(X_n: \ZZZ, \delta(n))}
{{\sf SA}(X_n, \delta(n))}.$$

\end{definition}

The effectiveness represents the ratio of how much portion of semantics of original message $\XXX$ is understood and conducted correctly by ${\sf R}$.

In the above-mentioned experiment, if robot $R$ is very capable, $R$ should correctly recognize various requests from $S$ and act accordingly as requested.
For example, $R$ recognizes a million ($\approx 2^{20}$) requests and behaves correctly as requested.
In this case, we can consider that the effectiveness complexity is approximately 20 bits (similar to 20 bits of a semantic channel rate).
On the other hand, if $R$ is not so capable,
for example, $R$ recognizes only 8 requests and behaves correctly, then its effectiveness complexity is only 3 bits. 

In this example, the requests by $S$ are formalized as elements of semantic information space $\mathbb{M}^{\overline{C}}$, and 
$(R,D)$ is considered as a functionality similar to a semantic channel in Section \ref{sec:s-channel-coding}, where various semantic errors occur. 
Hence, the capability of robot $R$  (effectiveness complexity) with a distribution of instructions, and the error correcting functionality in the system are characterized by the
%similar to the capacity of a semantic channel in Section \ref{sec:s-channel-coding}. 
%As such a notion, we introduce 
{\it effectiveness capacity} of $(R,D)$ 
and {\it effectiveness rate} of a semantic coding, which are defined in a manner similar to that of semantic channel capacity and     
semantic communication rate (Definition  
\ref{def:scc-scr}), respectively.
Roughly, the effectiveness capacity indicates the rate of the maximum capability of robot $R$ (effectiveness complexity) and the effectiveness rate is the communication rate of semantic (error correcting) coding. 

First we introduce the notion of an {\it effectiveness coding system} in a manner similar to that of the semantic channel coding system. 

\begin{definition}\label{def:ecs} (Effectiveness Coding System)

Effectiveness coding system 
$\ECS \set 
\ECS({\sf S}^{(\overline{C},u,{\sf code})},{\sf R}, {\sf D}, {\sf E};\XXX,\YYY,\ZZZ)$ 
consists of sender ${\sf S}$,
receiver ${\sf R}$, device ${\sf D}$
and evaluator ${\sf E}$.
Here, ${\sf S}$ has a coding machine, 
$(\overline{C},u,{\sf code})$,
and sends 
a sequential family of distributions, $\XXX$
to ${\sf R}$. Receiver ${\sf R}$ then 
performs actions given $\XXX$, and   
${\sf R}$'s conduct is observed by device ${\sf D}$, which outputs a sequential family of distributions, $\ZZZ$, and sends it to ${\sf E}$.
Evaluator ${\sf E}$ 
is a conditional oc-circuit 
under $\ZZZ$
and outputs $\YYY$.

Here, $\XXX$ is generated by 
an oc-circuit, ${\CCC} := 
(\overline{C}, u, \infty,  
\vec{m}_\infty)$, 
where
$\overline{C} \set 
({C}, N_u, N_s, N_m, N_r,L_y,s_1)$ 
(logic),   
$u \in \{0,1\}^{N_u}$
(universe space), 
$\vec{m}_\infty \in \mathbb{M}^{\overline{C}} \set 
\{ (m_i)_{i=1,2,..} \ \mid \  
m_i \in \{0,1\}^{N_m} \}$
(semantic information space),
and  
$\mathbb{M}^{\overline{C}}_n 
= \{0,1\}^{N_m K_n}$ 
(the $N_m K_n$-bit prefix of $\mathbb{M}^{\overline{C}}$ for $n$-bit output) 
($K_n \set \lceil n/L_y \rceil$).

For parameter $n \in \N$,
given  
$\vec{m}^+_{n} \in 
\{0,1\}^{k(n)}$, 
${\sf S}$ computes
$\vec{m}_{n} \set {\sf code}_n(\vec{m}^{+}_{n}) \in \mathbb{M}^{\overline{C}}_n$
and
$X_n 
\rset 
(\overline{C}, u, n, \vec{m}_{n})$,
where
$X_n \in \XXX$ is a distribution 
over $\{0,1\}^n$.
$X_n$ is input to receiver $\sf R$,
and 
$\sf R$'s conduct is observed by device ${\sf D}$, which  
outputs $Z_{\ell(n)} \in \ZZZ$,
where $Z_{\ell(n)}$ is a distribution 
over $\{0,1\}^{\ell(n)}$.
Evaluator ${\sf E}$ (conditional oc-circuit under $\ZZZ$) receives $Z_{\ell(n)}$ 
and outputs $Y_n \in \YYY$,
where $Y_n$ is a distribution over
$\{0,1\}^n$.

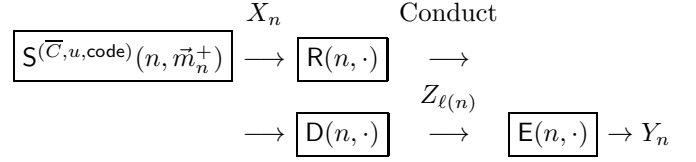
\begin{center}
\begin{tabular}{cccccccc}
 & $X_n$ & & 
 {\rm Conduct} & &
  &  \\
 \fbox{${\sf S}^{(\overline{C},u, 
{\sf code})}(n,\vec{m}^+_n)$} 
& $\longrightarrow$ & 
 \fbox{${\sf R}(n,\cdot)$} & 
 $\longrightarrow$  & 
 &
  & 
  \\
&  & &
 $Z_{\ell(n)}$ &  \\
& $\longrightarrow$ & 
 \fbox{${\sf D}(n,\cdot)$} &
 $\longrightarrow$ & 
  \fbox{${\sf E}(n,\cdot)$} 
    $\ra Y_n$\\

\end{tabular}
\end{center}

We say that 
$\ECS \set 
\ECS({\sf S}^{(\overline{C},u,{\sf code})},{\sf R}, {\sf D}, {\sf E};\XXX,\YYY,\ZZZ)$
correctly 
codes at precision level $\delta(\cdot)$ 
if there exists $n_0\in \mathbb{N}$ 
such that
for all $n \geq n_0$, \ 
$X_n \overset{\delta(n)}{\approx} Y_n$.

\end{definition}

\begin{definition}\label{def:ec-er} 
(Effectiveness Capacity and 
Effectiveness Rate)

Let 
$\ES \set ({\sf S}^{(\overline{C},u)}, {\sf R}, {\sf D};\XXX,\ZZZ)$
be a ``effectiveness system'' of effectiveness coding system 
$\ECS({\sf S}^{(\overline{C},u,{\sf code})},{\sf R}, {\sf D}, {\sf E};\XXX,\YYY,\ZZZ)$.
%where $\overline{C} \set (C, N_u, N_s, N_m, N_r, L_x, s_1)$, $u$, and $\mathbb{M}^{\overline{C}}$ are the logic, universe, and semantic information space of an oc-circuit for a family of distributions, respectively. 
Given $\vec{m}_\infty \in \mathbb{M}^{\overline{C}}$,
${\sf S}$ in $\ES$ computes
$\XXX \set \{X(\vec{m}_n)
\set X_n \rset (\overline{C}, u,n,\vec{m}_n) \}_{n \in \N}$, i.e., $\XXX \rset (\overline{C}, u, \infty, \vec{m}_\infty)$, where $\vec{m}_n$ is the 
$N_m \cdot \lceil n/L_y \rceil$-bit
prefix of 
$\vec{m}_\infty$ for $n$-bit output. 
Then, $\XXX \set \{X(\vec{m}_n)\}_{n \in \N}$ is input to $\sf R$ in $\ES$,
and 
$\sf D$ 
outputs $\ZZZ \set \{Z_{\ell(n)}\}_{n \in \N}$.

If 
for any $\vec{m}_\infty \in \mathbb{M}^{\overline{C}}$, 
$\XXX \rset (\overline{C}, u, \infty, \vec{m}_\infty)$ is a conditional semantic information source under $\ZZZ$ in effectiveness system $\ES$, 
we call $\ES$ ``{\it normal}''.

Let $\MMM \set \{M_n\}_{n \in \N}$ be a sequential family of distributions 
over $\mathbb{M}^{\overline{C}}$,
where
$M_n$ is a distribution 
over 
$\mathbb{M}^{\overline{C}}_n$,  
$n\in\N$, i.e., 
$M_n := \{ (\vec{m}_n, p_{\vec{m}_n}) 
\mid \vec{m}_n \in \mathbb{M}^{\overline{C}}_n\}$,
(see Definition \ref{def:seq-fd}
for the sequential family of distributions).

When effectiveness system $\ES$ is normal,  
``effectiveness  capacity'' ${\sf EC}$ of  
$\ES$ 
for $n \in \N$
(say $\ES_n$)
is
\begin{eqnarray}
&&
{\sf EC}(\ES_n,
\delta(n)) 
\set 
\nonumber
\\
&& \
\frac{1}{n}\cdot 
\max_{M_n \in \MMM^{\sf seq}_n}\{
H(M_n) 
- \ {\sf E}_{M_n} 
(\SA(X(\vec{m}_n): \ZZZ, \delta(n))
\},\ \ \ \ \
\label{eq:ec}
\end{eqnarray}
where 
$\MMM^{\sf seq}_n$ 
is the class of 
the $N_m \cdot \lceil n/L_y \rceil$-bit
prefix of sequential families of distributions,
i.e., $M_n \in \MMM^{\sf seq}_n$
is the $N_m \cdot \lceil n/L_y \rceil$-bit
prefix of a sequential family of distributions,
$H(\cdot)$ is the Shannon entropy and 
${\sf E}_{M_n}  
(\cdot)$
is the expectation value over the distribution of
$\vec{m}_n \rset M_n \in \MMM^{\sf seq}_n$.

``Effectiveness rate'' ${\sf ER}$ of 
effectiveness coding $(\overline{C},$ $u, {\sf code})$ 
in effectiveness coding system
$\ECS({\sf S}^{(\overline{C}, u,{\sf code})},$ ${\sf R}, {\sf D}, {\sf E};\XXX,\YYY,\ZZZ)$ for $n \in \N$ is
\begin{eqnarray}
{\sf ER}({\sf code}_n) \set 
k(n)/n. \ \ \  
\end{eqnarray}

We say a effectiveness coding system, $\ECS$, is ``normal,'' if the effectiveness system, $\ES$, of $\ECS$ is normal.

\end{definition}

We can also 
define the ``uniformity'' of 
normal effectiveness coding system $\ECS$
in the same manner as that for 
the normal semantic channel coding system
described in Definition \ref{def:usc}.

\begin{theorem}\label{thm:ec}
 (Effectiveness Coding Theorem)

There exists a uniform effectiveness coding system, 
$\ECS$,
that correctly codes 
at precision level $\delta(\cdot)$
and 
for any $\epsilon \ (0 <\epsilon)$, 
there exits $n_0\in\N$ such that 
for all $n > n_0$, \ 
\begin{eqnarray*}
&&
{\sf EC}(\ES_n,\delta(n)) - \epsilon 
< 
{\sf ER}({\sf code}_n) 
<
{\sf EC}(\ES_n,\delta(n)). 
\end{eqnarray*}

There exists no normal effectiveness coding system, 
$\ECS$,
that correctly 
codes at precision level $\delta(\cdot)$
(with a negligible error probability) 
and
for any $\epsilon \ (0 <\epsilon)$,
there exits $n_0\in\N$ such that 
for all $n > n_0$, \  
$$ 
{\sf EC}(\ES_n,
\delta(n)) < {\sf ER}({\sf code}_n). 
$$

\end{theorem}

\section{Conclusion}

Approximately seven decades have passed since Warren Weaver published his two insightful and prescient articles \cite{Weaver48,Weaver49} that clearly indicated two research directions in science, organized complexity and semantic information theory. Although the articles stimulated and encouraged these research areas, it is hard to say that these areas have been well established in science, and Weaver would be disappointed to know it.

Moreover, he might be disappointed to learn 
that no %research 
study has been done on the relation and integration of these areas, since he could have realized the relationship between the areas considering that these articles were written at almost the same time. 

The aim of this paper is to pursue the research directions that Weaver indicated. 
%and to found the {\it organized} %complexity version of Shannon theory, %while Shannon theory was 
%%accomplished 
%constructed on {\it disorganized} %complexity.     
%
This paper first quantitatively defined the organized complexity.  
%as Shannon %quantitatively defined 
%did the disorganized complexity of %information as entropy for his (syntactic) information theory.
The proposed definition for the first time   simultaneously captures the three major features of organized complexity and satisfies all of the criteria for organized complexity measures introduced in this paper. 
We then applied the organized complexity measure to develop our semantic information theory, where we presented the first formal definition of a semantic information amount
that is based only on concretely defined notions,
and unveil several fundamental properties in the semantic information theory.
Through this organized complexity measure, we offered a unified paradigm of organized complexity and semantic information theory.

%As suggested by %described in the article of Weaver \cite{Weaver48}, 
Organized complexity is an interdisciplinary concept straddling physics, cosmology, biology, ecology, sociology, and informatics. Thus, the proposed organized complexity measure could be a core notion in such interdisciplinary areas, and for example, offer some basis for tackling the problems posed in \cite{LinDavRus13}. %such as ``What is complexity,'' ``Is complexity increasing?'' and ``Does it have an ultimate limit?''        

%The proposed organized complexity measure has pushed toward a new horizon of the interdisciplinary field, organized complexity.
%%initiated a new frontier, 
%integrated the areas of organized complexity aynd semantic information theory. 
%An undeveloped fertile land would be expected at this frontier and plowed fertile land to allow new seeds to grow in this new frontier.
%a fertile land would remain undeveloped %at this frontier. 

\end{document}